\documentclass[journal]{IEEEtran}
\ifCLASSINFOpdf
\else
\fi
\usepackage{cite}
\usepackage{amsmath}
\usepackage[pdftex]{graphicx} % support the \includegraphics command and options
\usepackage{subfig}
\usepackage{array} % for better arrays \left(eg matrices\right) in maths
\usepackage{fixltx2e}

\usepackage{booktabs} % for much better looking tables
\usepackage{amssymb}
\usepackage{mathtools}
\usepackage{cancel} % For \cancelto
\usepackage{bm} % for \bs
\usepackage{esint} % for contour integrals etc.
\usepackage{amsthm}
\usepackage{tikz}
\usepackage{tikz-3dplot}
\usetikzlibrary{backgrounds,3d,shadings,shapes.misc,decorations.pathmorphing,shapes,calc,fadings,shadows.blur}
\usepackage{amssymb,amsmath}

\usepackage{bm}
\newcommand{\bs}[1]{\bm{\mathrm{#1}}}

\newcommand{\vect}[1]{\vec{#1}}
\newcommand{\nhat}{\ensuremath{\hat{n}}}
\newcommand{\rhat}{\ensuremath{\hat{r}}}

\newcommand{\vecprime}[1]{\vect{#1}^{\,\prime}}

\renewcommand{\r}{\left(\vect{r}\right)}
\newcommand{\rp}{\left(\vecprime{r}\right)}
\newcommand{\rrp}{\left(\vect{r}, \vecprime{r}\right)}

\newcommand{\depk}[1][]{\left(k_{#1}\right)}

\newcommand{\rrpk}[1][]{\left(k_{#1}, \vect{r}, \vecprime{r}\right)}
\newcommand{\rrpki}[1][]{\left(k_i, \vect{r}, \vecprime{r}\right)}

\newcommand{\dyadic}[1]{\overline{\overline{\mathbf{#1}}}}
\newcommand{\matr}[1]{\bs{#1}}

\newcommand{\abs}[1]{\left \lvert #1 \right\rvert }
 
\newcommand{\junk}[1] {}

\def\XXint#1#2#3{{\setbox0=\hbox{$#1{#2#3}{\int}$}
\vcenter{\hbox{$#2#3$}}\kern-.5\wd0}}

\newcommand*\widebar[1]{%
  \hbox{%
    \vbox{%
      \hrule height 0.5pt % The actual bar
      \kern0.3ex%         % Distance between bar and symbol
      \hbox{%
        \kern-0.05em%      % Shortening on the left side
        \ensuremath{#1}%
        \kern-0.05em%      % Shortening on the right side
      }%
    }%
  }%
} 

\renewcommand{\epsilon}{\varepsilon}

\newcommand{\opL}{\ensuremath{\mathcal{L}}} % L operator
\newcommand{\opK}{\ensuremath{\mathcal{K}}} % K operator

\newcommand{\Lmat}[1][]{{\matr{L}_{#1}}}
\newcommand{\LmatA}[1][]{{\matr{L}^{(A)}_{#1}}}
\newcommand{\LmatPhi}[1][]{{\matr{L}^{(\phi)}_{#1}}}
\newcommand{\Kmat}[1][]{{\matr{K}_{#1}}}

\newcommand{\figref}[1]{Fig.~\ref{#1}}
\newcommand{\secref}[1]{Section~\ref{#1}}
\newcommand{\tabref}[1]{Table~\ref{#1}}

\usepackage{tcolorbox}

\usepackage{textpos}

\newcommand{\mySubtitle}[1]%
{%
	\begin{textblock}{14.0}(0.7, 2.9)
		\textbf{#1}%
	\end{textblock}%
}%

\newcommand{\green}[1]{\textcolor{green!0!black}{#1}}

\newcommand{\Ecolor}[1]{{#1}}
\newcommand{\Hcolor}[1]{{#1}}
\newcommand{\Jcolor}[1]{{#1}}

\newcommand{\Er}[1][]{\Ecolor{\vect{E}_{#1}\r}}

\newcommand{\Emat}[1][]{\Ecolor{\matr{E}_{#1}}}

\newcommand{\Hr}[1][]{\Hcolor{\vect{H}_{#1}\r}}

\newcommand{\Hmat}[1][]{\Hcolor{\matr{H}_{#1}}}

\newcommand{\Jr}[1][]{\ensuremath{\Jcolor{\vect{J}_{#1}\r}}}
\newcommand{\Jrp}[1][]{\ensuremath{\Jcolor{\vect{J}_{#1}\rp}}}

\newcommand{\Jmat}[1][]{\ensuremath{\Jcolor{\matr{J}_{#1}}}}

\newcommand{\Grrp}[1][]{\ensuremath{\green{G_{#1}\rrp}}}

\newcommand{\Grrpko}[1][]{\ensuremath{\green{G_{#1}\rrpk[0]}}}
\newcommand{\Grrpkom}[1][]{\ensuremath{\green{G_{#1}\rrpk[0]}}}
\newcommand{\Grrpm}[1][]{\ensuremath{\green{G_{#1}\rrp}}}

\newcommand{\Grrpki}[1][]{\ensuremath{\green{G\rrpki}}}

\newcommand{\Gdyadicrrpko}[1][]{\ensuremath{\green{\dyadic{G}_{#1}\rrpk[0]}}}

\newcommand{\Ktheta}[2][]{\ensuremath{\green{K_{#1}(\theta_{#2})}}}
\newcommand{\tKtheta}[2][]{\ensuremath{\green{\widetilde{K}_{#1}(\theta_{#2})}}}

\newcommand{\Gtheta}[2][]{\ensuremath{\green{G_{#1}(\theta_{#2})}}}
\makeatletter
\newlength\numerator@height
\newlength\frac@height
\newsavebox\numerator@box
\newsavebox\frac@box

\newcommand\dfracparens[3]{%
	\sbox{\numerator@box}{\ensuremath{#1}}%
	\sbox{\frac@box}{\ensuremath{\dfrac{#1}{#2}}}%
	\settoheight{\frac@height}{\usebox{\frac@box}}%
	\settoheight{\numerator@height}{\usebox{\numerator@box}}%
	\addtolength{\frac@height}{-\numerator@height}%
	\usebox{\frac@box}%
	\raisebox{\frac@height}{%
		\( \left( {#3} \right)
		\)}%
}
\makeatother

\usepackage{enumitem}

\begin{document}
% \title{A Fast Frequency Sweep Technique Based on the Adaptive Integral Method for Electromagnetic Modeling with Surface Integral Equations}
%\title{AIMx: An Extended Adaptive Integral Method for~Fast~Frequency~Sweeps in Surface Integral Equation-Based Electromagnetic Modeling}
%\title{AIMx: A Fast Frequency Sweep Algorithm for~Electromagnetic~Modeling with Surface~Integral~Equations}
%\title{AIMx: A Fast Frequency Sweep Technique Based on Kernel Decomposition for Electromagnetic Modeling with Surface~Integral~Equations}
\title{AIMx: An Extended Adaptive Integral Method for~the~Fast~Electromagnetic~Modeling of~Complex~Structures}
%\title{AIMx: An Extended Adaptive Integral Method for~Fast Broadband Electromagnetic Modeling}
%\title{AIMx:~An~Extended~Adaptive~Integral~Method Based~on~Kernel~Decomposition for~Fast~Electromagnetic~Modeling}

\author{Shashwat~Sharma,~\IEEEmembership{Graduate Student Member,~IEEE,}
        and~Piero~Triverio,~\IEEEmembership{Senior Member,~IEEE}% <-this % stops a space
\thanks{Manuscript received $\ldots$; revised $\ldots$.}
\thanks{S. Sharma and P. Triverio are with the Edward S. Rogers Sr. Department of Electrical \& Computer Engineering, University of Toronto, Toronto,
ON, M5S 3G4 Canada, e-mails: shash.sharma@mail.utoronto.ca, piero.triverio@utoronto.ca.}% <-this % stops a space
\thanks{This work was supported by the Natural Sciences and Engineering Research 
	Council of Canada (Collaborative Research and Development Grants 
	program), by Advanced Micro Devices, and by CMC Microsystems.}}


\markboth{IEEE Transactions on Antennas and Propagation}%
{Sharma \MakeLowercase{\textit{et al.}}: Bare Demo of IEEEtran.cls for IEEE Journals}



\maketitle

\begin{abstract}
Surface integral equation (SIE) methods are of great interest for the efficient electromagnetic modeling of various devices, from integrated circuits to antenna arrays.
Existing acceleration algorithms for SIEs, such as the adaptive integral method (AIM), enable the fast approximation of interactions between well-separated mesh elements.
Nearby interactions involve the singularity of the kernel, and must instead be computed accurately with direct integration at each frequency of interest, which can be computationally expensive.
We propose a novel algorithm for reducing the cost-per-frequency of near-region computations for both homogeneous and layered background media.
In the proposed extended AIM (AIMx), the SIE operators are decomposed into a frequency-independent term containing the singularity of the kernel, and a nonsingular frequency-dependent term.
Direct integration is only required for the frequency-independent term, and can be reused at each frequency, leading to significantly faster frequency sweeps.
The frequency-dependent term is captured with good accuracy via fast Fourier transform (FFT)-based acceleration even in the near region, as confirmed with an error analysis.
The accuracy and efficiency of the proposed method are demonstrated through numerical examples drawn from several applications, and CPU times are significantly reduced by factors ranging from three to 16.
\end{abstract}

\begin{IEEEkeywords}
Electromagnetic modeling, surface integral equations, adaptive integral method, fast frequency sweep.
\end{IEEEkeywords}

% \fi
\IEEEpeerreviewmaketitle

\section{Introduction}

\IEEEPARstart{F}{ull}-wave electromagnetic modeling is crucial in many design workflows, ranging from the characterization of on-chip interconnect networks, to the prediction and synthesis of antenna radiation profiles.
The growing complexity of modern electronic devices places a commensurate demand on full-wave simulation tools, especially since device characterization is often required over a wide frequency band.

Volumetric methods such as the finite element method (FEM)~\cite{FEMJin} and volume integral equations (VIEs)~\cite{PEEC01,fastmaxwell,VolIE01} have proved to be robust, but require a volumetric discretization of conductive and dielectric objects.
The FEM additionally requires a discretization of the space in between objects.
These requirements lead to systems of equations with a large number of unknowns.
In contrast, surface integral equation (SIE)-based techniques require only a discretization of the surface of conductive and dielectric objects~\cite{PMCHWT02,gibcHmatDanJiao,aefie2,eaefie01,DSA08,AWPLSLIM}.
However, SIE methods lead to a dense system matrix whose solution is computationlly prohibitive for large problems.
Techniques based on SIEs heavily rely on various acceleration algorithms~\cite{FMAorig,enghetaFMA,MLFMA,MFFMA,AIMbles}, where the system of equations is solved iteratively. The products between a system matrix $\matr{A}$ and a vector $\matr{x}$ are computed at each iteration as
\begin{align}
	\matr{A}\matr{x} = \matr{A}_{\mathrm{NR}}\matr{x} + \matr{A}_{\mathrm{FR}}\matr{x}.\label{eq:acc}
\end{align}
where $\matr{A}_{\mathrm{NR}}$ and $\matr{A}_{\mathrm{FR}}$ contain entries of $\matr{A}$ corresponding to nearby (``near-region'') and well-separated (``far-region'') mesh elements, respectively.
Both $\matr{A}_{\mathrm{NR}}$ and $\matr{A}_{\mathrm{FR}}$ depend on the Green's function of the surrounding medium.
The entries of $\matr{A}_{\mathrm{NR}}$ are dominant, and must be computed accurately with direct integration.
Far-region interactions in $\matr{A}_{\mathrm{FR}}$ are weaker, therefore the product $\matr{A}_{\mathrm{FR}}\matr{x}$ can be computed in a fast but approximate way, without having to assemble $\matr{A}_{\mathrm{FR}}$.

The fast multipole method (FMM)~\cite{FMAorig,enghetaFMA} and its multi-level and mixed-form extensions~\cite{MLFMA,MFFMA} are popular acceleration algorithms.
These techniques were developed specifically for the homogeneous medium Green's function.
However, many practical applications involve objects embedded in stratified dielectric media, which may require the more complicated multilayer Green's function (MGF)~\cite{MGF01}.
The extension of FMM-based techniques to handle the MGF is not straightforward, and requires either multiple complex exponential terms~\cite{MLFMADCIM} or numerical integration in the complex plane~\cite{FIPWA}.

Another class of acceleration algorithms arises from the use of an auxiliary coarse grid to compute $\matr{A}_{\mathrm{FR}}\matr{x}$ quickly.
The adaptive integral method (AIM)~\cite{AIMbles} involves superimposing a regular grid on the structure, and projecting currents and charges from the original mesh onto the auxiliary grid.
These grid currents and charges behave as effective sources, and are chosen via moment-matching to generate approximately the same fields in the far region as the original sources on the mesh.
The translation invariance of the Green's function is exploited by using Fast Fourier Transforms (FFTs) to accelerate the computation of $\matr{A}_{\mathrm{FR}}\matr{x}$.
An advantage of the AIM is its kernel-independence, which makes it relatively easy to incorporate the MGF~\cite{okh_AIM_MGF_3D,AIM_MGF_2D,convcorr_AIM,CPMT2019arxiv}.
A closely related technique, the precorrected FFT (pFFT) method~\cite{pfftmain}, also makes use of an auxiliary grid along with FFT-based acceleration, and differs from the AIM in the way the effective grid sources are calculated.
Other related acceleration techniques include the sparse-matrix/canonical grid (SMCG) method~\cite{smcg}, the IE-FFT algorithm~\cite{iefft}, and several others~\cite{gifft,mlgfim,mlgffft}.

In existing acceleration methods, generating $\matr{A}_{\mathrm{NR}}$ with direct integration may still have a significant cost, particularly for densely-packed structures with fine features, which often require an extremely fine local mesh.
This cost is further increased in the presence of layered media, since the MGF is significantly more expensive to compute than the homogeneous Green's function.
Moreover, since the Green's function is singular when the source and observation points coincide, the entries of $\matr{A}_{\mathrm{NR}}$ must be computed carefully with advanced singularity extraction or cancellation routines~\cite{gibson}, which further add to the computational cost.
Finally, the computation of $\matr{A}_{\mathrm{NR}}$ must be repeated at each frequency of interest.
Methods to interpolate simulation results across frequency have been proposed~\cite{awe,ZutterAFS,MeyerASA,AntoniniMacromodPEEC1,AntoniniMacromodPEEC2}, but still require discrete simulation data at multiple frequencies to gather sufficient sample points.
This is particularly onerous for structures whose terminal response varies sharply with frequency, for example due to resonances.

In this contribution, we propose a novel acceleration algorithm based on the AIM, which requires computing $\matr{A}_{\mathrm{NR}}$ only once for the entire frequency sweep.
In the proposed extended adaptive integral method (AIMx), the kernel of each integral operator is carefully decomposed into a frequency-independent term which contains the singularity of the kernel, and a frequency-dependent term which is continuous and nonsingular everywhere in space.
This leads to a convenient decomposition of the associated system matrix into frequency-independent and frequency-dependent parts.
Through an error analysis, we demonstrate that the frequency-dependent part of the system matrix can be computed with good accuracy with FFT-based acceleration, even in the near region.
Therefore, matrix-vector products involving the frequency-dependent part of each matrix operator are accelerated with the use of FFTs.
The frequency-\emph{independent} part of the matrix is expressed as in \eqref{eq:acc}, where the near-region entries are computed accurately with direct integration.
As a consequence, the key advantage of the proposed method is that direct integration needs to be performed only once at the beginning of a frequency sweep, and then can be reused at each frequency point.
We describe in detail the proposed matrix decomposition for both the single- and double-layer potential matrix operators~\cite{book:colton}, for both homogeneous and layered background media.
This leads to significantly faster frequency sweeps for a wide range of problems requiring full-wave electromagnetic characterization.

A similar concept was also exploited in~\cite{ChewAEFIEperturbOrig,ChewAEFIEperturb}, where the Taylor expansion of the Green's function was used to obtain a series of matrices with a multiplicative frequency dependence, to accurately handle very low frequencies.
However, the method in~\cite{ChewAEFIEperturbOrig} applies to low-frequency problems only, while the one in~\cite{ChewAEFIEperturb} requires using an increasingly large number of Taylor series terms, and a correspondingly large number of matrix-vector products, as frequency increases.
Therefore, both CPU time and memory cost in~\cite{ChewAEFIEperturb} increase with frequency.
Unlike~\cite{ChewAEFIEperturbOrig,ChewAEFIEperturb}, the proposed method does not explicitly use a series expansion, and therefore avoids the need for an increasingly large number of matrix operators at high frequencies.
This key distinction makes the proposed method well suited for both low- and high-frequency problems.
Furthermore, the proposed method is applicable to both the single- and double-layer potential operators~\cite{book:colton} for both the free space and the multilayer Green's function.
Unlike the methods referenced above, the extension to layered media is simple in the proposed technique, since it does not require an explicit Taylor expansion of the components of the dyadic Green's function.

The Taylor expansion of the Green's function has also been used to speed up near-region computations in\cite{taylor1} for free space, and~\cite{konno,EPEPS2018} for layered media.
However, unlike these methods, the proposed technique does not explicitly utilize a Taylor expansion, and allows the direct integration step to be performed independently of the frequency.

The paper is organized as follows: \secref{sec:background} summarizes the SIE formulation considered and the conventional AIM. \secref{sec:methods} provides a detailed description of the proposed method, and \secref{sec:validation} shows analytical and numerical error analyses to support the hypothesis that frequency-dependent terms can be interpolated accurately in the near region.
In \secref{sec:results}, we demonstrate the accuracy and efficiency of the proposed method compared to the conventional AIM for a series of numerical examples, including canonical geometries, on-chip devices and large antenna arrays.
A discussion of the advantages of the proposed method is given in \secref{sec:discussion}, followed by concluding remarks in \secref{sec:conclusion}.

\section{Background}\label{sec:background}

We consider time-harmonic fields with a time dependence of $e^{j\omega t}$.
Vector quantities are written with an overhead arrow, for example $\vect{a}\r$.
Primed and unprimed coordinates represent source and observation points, respectively.
Matrices and column vectors are written in bold letters, such as $\matr{A}$ and $\matr{x}$.
Dyadic quantities are denoted with double overbars, as in $\dyadic{A}$.

\subsection{Formulation}\label{sec:methods:formulation}

We consider the problem of electromagnetic scattering from a perfect electric conductor (PEC) object to illustrate the proposed method, though it is also applicable to penetrable objects.
The surface of the PEC object is denoted as $\mathcal{S}$, with outward unit normal $\nhat$.
The conductor is embedded in free space, denoted by $\mathcal{V}_0$.
The layered medium case will be considered in \secref{sec:methods:aimx:mgf}.
The unknown induced electric surface current density $\Jr$ on $\mathcal{S}$ due to an incident field, $[\Er[\mathrm{inc}], \Hr[\mathrm{inc}]]$, $\vect{r} \in \mathcal{V}_0$, satisfies the electric field integral equation (EFIE)~\cite{gibson},
\begin{align}
  -j\omega\mu_0\, \nhat \times \opL \bigl[\Jrp\bigr]\depk[0] = \nhat \times \Er[\mathrm{inc}],\label{eq:EFIEout}
\end{align}
where~$\opL$ is the single-layer potential operator~\cite{book:colton,gibson},
\begin{multline}
  \opL\bigl[\vect{X}\rp\bigr]\depk[0] \\= \int_\mathcal{S} \left( \vect{X}\rp + \frac{1}{k_0^2}\nabla'\nabla' \cdot \vect{X}\rp \right) \Grrpko\, d\mathcal{S}.\label{eq:opL}
\end{multline}
In \eqref{eq:EFIEout} and \eqref{eq:opL}, $\omega$ is the cyclical frequency, $\mu_0$ and $k_0$ are the permeability and wave number of free space, respectively, and the kernel $\Grrpko$ is the free space Green's function,
\begin{align}
  \Grrpko = \frac{e^{-j k_0 r}}{4\pi r},\label{eq:hgf}
\end{align}
where $r = |\vect{r} - \vect{r}^{\,'}|$.

Next, a triangular mesh is generated for $\mathcal{S}$, and \eqref{eq:EFIEout} is discretized with the method of moments (MoM).
The electric surface current density $\Jr$ is expanded with Rao-Wilton-Glisson (RWG) basis functions~\cite{RWG} defined on pairs of adjacent triangles, and \eqref{eq:EFIEout} is tested with $\nhat\times\text{RWG}$ basis functions to get the matrix equation
\begin{align}
  -j\omega\mu_0 \left(\LmatA + k_0^{-2}\LmatPhi\right) \Jmat = \Emat[\mathrm{inc}],\label{eq:EFIEdis}
\end{align}
where $\LmatA$ and $\LmatPhi$ are, respectively, the vector and scalar potential parts of the discretized ${\nhat\times\opL}$ operator.
Column vectors $\Jmat$ and $\Emat[\mathrm{inc}]$ contain coefficients of the basis functions associated with $\Jr$ and $\nhat\times\Er[\mathrm{inc}]$, respectively.
The hypersingular part of \eqref{eq:opL} is handled by transferring the gradient operator $\nabla'$ to the testing function~\cite{gibson}.

\subsection{The Conventional AIM}\label{sec:methods:aim}

In the AIM, a regular grid is superimposed on the bounding box associated with $\mathcal{S}$, and $\Lmat$ is written as
\begin{align}
  \Lmat(k_0) = \Lmat[\mathrm{NR}](k_0) + \Lmat[\mathrm{FR}](k_0),\label{eq:aimL}
\end{align}
where $\Lmat$ represents either $\LmatA$ or $\LmatPhi$. Matrix $\Lmat[\mathrm{NR}](k_0)$ is sparse and contains the near-region entries of $\Lmat(k_0)$, and is computed using direct integration.
Using direct integration to compute $\Lmat[\mathrm{NR}](k_0)$ is necessary to accurately capture the singularity of $\Grrpko$ when $\vect{r} = \vecprime{r}$.
Matrix $\Lmat[\mathrm{FR}](k_0)$ contains far-region interactions, which are less dominant and are approximated as
\begin{align}
  \Lmat[\mathrm{FR}](k_0) \approx \matr{W}^{(\Lmat)}\matr{H}(k_0)\matr{P}^{(\Lmat)} - \Lmat[\mathrm{P}](k_0),\label{eq:aimLFR}
\end{align}
where matrix $\matr{P}^{(\Lmat)}$ projects sources from the mesh to the AIM grid, $\matr{H}(k_0)$ is the convolution matrix encoding the Green's function, and $\matr{W}^{(\Lmat)}$ interpolates the computed potentials back onto the mesh.
Due to the Toeplitz nature of $\matr{H}(k_0)$, matrix-vector products involving $\Lmat[\mathrm{FR}](k_0)$ are accelerated with FFTs~\cite{AIMbles,pfftmain}.
The FFTs involved in the computation of $\matr{W}^{(\Lmat)}\matr{H}(k_0)\matr{P}^{(\Lmat)}$ operate on the entire AIM grid, including both the near and far regions of a given source.
Therefore, to avoid double-counting the near-region matrix entries, the precorrection matrix $\Lmat[\mathrm{P}](k_0)$ is used to cancel out the contribution of $\matr{W}^{(\Lmat)}\matr{H}(k_0)\matr{P}^{(\Lmat)}$ in the near region~\cite{AIMbles,pfftmain}.
Matrix $\Lmat[\mathrm{P}](k_0)$ is defined as
\begin{align}
  \Lmat[\mathrm{P}](k_0) = \matr{W}^{(\Lmat)}\matr{H}_\mathrm{NR}(k_0)\matr{P}^{(\Lmat)},\label{eq:aimLP}
\end{align}
where $\matr{H}_\mathrm{NR}(k_0)$ is the convolution matrix associated with the near region.

Though $\Lmat[\mathrm{NR}](k_0)$ is sparse, its assembly with direct integration is expensive for large problems, particularly when the structure is densely packed and requires a fine mesh to resolve intricate features.
This is especially cumbersome for layered background media which require the multilayer Green's function (MGF)~\cite{MGF01}, which is significantly more expensive to compute than its free space counterpart.
Both $\Lmat[\mathrm{NR}](k_0)$ and $\Lmat[\mathrm{P}](k_0)$ must be generated for each frequency, which can be a bottleneck for wideband frequency sweeps.

\section{Proposed Method}\label{sec:methods}

The conventional AIM has the following limitation: due to the singularity of the Green's function at ${r = 0}$, the entries of $\Lmat[\mathrm{NR}](k_0)$ cannot be approximated with interpolation, and must instead be computed with direct integration.
The singular behavior of $\Grrpko$ may be investigated further through a Taylor expansion about ${r = 0}$,
\begin{multline}
  \Grrpko = \frac{1}{4\pi r} \\+ \frac{\left(-jk_0\right)}{4\pi} + \frac{\left(-jk_0\right)^2}{4\pi\cdot 2!} r + \frac{\left(-jk_0\right)^3}{4\pi\cdot 3!} r^2 + \cdots.\label{eq:hgftaylor}
\end{multline}
The first term on the right-hand side of~\eqref{eq:hgftaylor} is frequency-independent, and is the static Green's function~\cite{book:colton}.
The remaining terms on the right-hand side of~\eqref{eq:hgftaylor} are frequency-dependent.
Therefore, we may write
\begin{align}
  \Grrpko = \Grrp[\mathrm{s}] + \Grrpko[\mathrm{d}],\label{eq:hgfdecomp0}
\end{align}
where
\begin{align}
  \Grrp[\mathrm{s}] = \frac{1}{4\pi r}\label{eq:hgfs}
\end{align}
is frequency-independent, and
\begin{align}
  \Grrpko[\mathrm{d}] = \frac{e^{-j k_0 r} - 1}{4\pi r}.\label{eq:hgfd}
\end{align}
is frequency-dependent.
Equations~\eqref{eq:hgftaylor} and~\eqref{eq:hgfdecomp0} reveal the key observations underlying the proposed method:
\begin{itemize}
  \item the singularity of the Green's function is contained entirely in its frequency-independent part, $\Grrp[\mathrm{s}]$;
  \item each frequency-\emph{dependent} term on the right-hand side of~\eqref{eq:hgftaylor} is either constant, or decays to~$0$ as ${r \rightarrow 0}$, and $\Grrpko[\mathrm{d}]$ is a continuous function for any finite~$r$.
\end{itemize}
This leads to our main hypothesis: $\Grrpko[\mathrm{d}]$ is amenable to approximation by interpolation, even in the near region where~$r$ is small.
Therefore, $\Lmat[\mathrm{NR}](k_0)$ may also be decomposed into frequency-independent and frequency-dependent parts.
Then, only the frequency-independent part needs to be computed accurately via direct integration, which does not have to be repeated at each frequency point.
Instead, the frequency-dependent part is computed in an accelerated manner with FFT-based acceleration \emph{even in the near region}, leading to the proposed extended adaptive integral method~(AIMx).

\subsection{AIMx for the Single-Layer Potential Operator}\label{sec:methods:L}

Using \eqref{eq:hgfdecomp0} in \eqref{eq:opL} allows decomposing the single-layer potential operator $\opL$ as
\begin{align}
  \opL\bigl[\vect{X}\rp\bigr]\depk[0] = \opL_{\mathrm{s}}\bigl[\vect{X}\rp\bigr] + \opL_{\mathrm{d}}\bigl[\vect{X}\rp\bigr]\depk[0],\label{eq:opLdecomp2}
\end{align}
where the frequency-independent part is
\begin{multline}
  \opL_{\mathrm{s}}\bigl[\vect{X}\rp\bigr] \\= \int_\mathcal{S} \left( \vect{X}\rp + \frac{1}{k_0^2}\nabla'\nabla' \cdot \vect{X}\rp \right) \Grrp[\mathrm{s}]\, d\mathcal{S},\label{eq:opLs}
\end{multline}
and the frequency-dependent part is
\begin{multline}
  \opL_{\mathrm{d}}\bigl[\vect{X}\rp\bigr]\depk[0] \\= \int_\mathcal{S} \left( \vect{X}\rp + \frac{1}{k_0^2}\nabla'\nabla' \cdot \vect{X}\rp \right) \Grrpko[\mathrm{d}]\, d\mathcal{S}.\label{eq:opLd}
\end{multline}
Correspondingly, $\Lmat$ may be decomposed as
\begin{align}
  \Lmat(k_0) = \Lmat[\mathrm{s}] + \Lmat[\mathrm{d}](k_0),\label{eq:Lmatdecomp}
\end{align}
where $\Lmat[\mathrm{s}]$ and $\Lmat[\mathrm{d}](k_0)$ represent either the vector or scalar potential parts of the discretized $\opL_{\mathrm{s}}$ and $\opL_{\mathrm{d}}$ operators, respectively.
The goal is to avoid the explicit assembly of $\Lmat[\mathrm{d}](k_0)$, and to instead directly compute the product of $\Lmat[\mathrm{d}](k_0)$ with a vector with the use of FFTs.

To this end, only the frequency-independent matrix $\Lmat[\mathrm{s}]$ is split into near- and far-region contributions as per~\eqref{eq:aimL},
\begin{align}
  \Lmat(k_0) = \Lmat[\mathrm{s,NR}] + \Lmat[\mathrm{s,FR}] + \Lmat[\mathrm{d}](k_0).\label{eq:Lmatdecomp2}
\end{align}
In~\eqref{eq:Lmatdecomp2}, the singularity of $\Grrpko$ is contained entirely in the sparse, frequency-independent matrix $\Lmat[\mathrm{s,NR}]$, which is computed accurately with direct integration.
The AIM is employed to express the remaining matrices,
\begin{multline}
  \Lmat[\mathrm{s,FR}] + \Lmat[\mathrm{d}](k_0) \approx -\Lmat[\mathrm{s,P}] + \matr{W}^{(\Lmat)}\matr{H}_{\mathrm{s}}\matr{P}^{(\Lmat)} \\+ \matr{W}^{(\Lmat)}\matr{H}_{\mathrm{d}}(k_0)\,\matr{P}^{(\Lmat)},\label{eq:Lmatdecomp3}
\end{multline}
where 
\begin{align}
  \Lmat[\mathrm{s,P}] = \matr{W}^{(\Lmat)}\matr{H}_\mathrm{s,NR}\matr{P}^{(\Lmat)},\label{eq:aimLPs}
\end{align}
is the near-region precorrection matrix associated to $\Lmat[\mathrm{s,NR}]$, and $\matr{H}_{\mathrm{s}}$ and $\matr{H}_{\mathrm{d}}(k_0)$ are convolution matrices depending on $\Grrp[\mathrm{s}]$ and $\Grrpko[\mathrm{d}]$, respectively.
Matrix $\matr{H}_\mathrm{s,NR}$ is the convolution matrix for near-region precorrections, and encodes $\Grrp[\mathrm{s}]$.
It should be noted that precorrections are required only for $\Lmat[\mathrm{s,NR}]$, because both near- \emph{and} far-region contributions of $\Lmat[\mathrm{d}](k_0)$ are approximated grid-based FFTs.
Since ${\matr{H}_{\mathrm{s}} + \matr{H}_{\mathrm{d}}(k_0) = \matr{H}(k_0)}$, \eqref{eq:Lmatdecomp3} can be simplified further by combining its last two terms,
\begin{align}
  \Lmat(k_0) \approx \Lmat[\mathrm{s,NR}] - \Lmat[\mathrm{s,P}] + \matr{W}^{(\Lmat)}\matr{H}(k_0)\matr{P}^{(\Lmat)}.\label{eq:LAIMx}
\end{align}
Equation~\eqref{eq:LAIMx} is the proposed AIMx matrix decomposition for the single-layer potential operator.
Finally, using~\eqref{eq:LAIMx} in~\eqref{eq:EFIEdis} yields the AIMx-accelerated EFIE for PECs,
\begin{multline}
  -j\omega\mu_0 \left[ \LmatA[\mathrm{s,NR}] - \LmatA[\mathrm{s,P}] + \matr{W}^{(A)}\matr{H}(k_0)\matr{P}^{(A)} \right. \\ \left. + \frac{1}{k_0^2}\left( \LmatPhi[\mathrm{s,NR}] - \LmatPhi[\mathrm{s,P}] + \matr{W}^{(\phi)}\matr{H}(k_0)\matr{P}^{(\phi)} \right) \right] \Jmat = \Emat[\mathrm{inc}],\label{eq:EFIEAIMx}
\end{multline}
where the superscript~$(A)$ or~$(\phi)$ indicates that the associated matrix corresponds to the vector or scalar potential part of~$\opL$, respectively.

In~\eqref{eq:EFIEAIMx}, all frequency-dependent effects are approximated with using the auxiliary grid, and the associated matrix-vector products are accelerated with FFTs.
The key advantage of the proposed method is that the entries of $\LmatA[\mathrm{s,NR}]$, $\LmatPhi[\mathrm{s,NR}]$, $\LmatA[\mathrm{s,P}]$ and $\LmatPhi[\mathrm{s,P}]$ are frequency-independent, and need only be computed once, enabling significantly faster frequency sweeps compared to the conventional AIM.
An important property of the proposed method is its simplicity of implementation.
In an existing implementation of the conventional AIM, the proposed method requires only two changes:
\begin{enumerate}[label=(\alph*)]
  \item the near-region and precorrection matrices, $\Lmat[\mathrm{s,NR}]$ and $\Lmat[\mathrm{s,P}]$, must be assembled using the kernel $\Grrp[s]$ rather than $\Grrpko$;
  \item a minor modification to $\matr{H}$ must be made, which is described in \secref{sec:methods:aimx:precorr}
\end{enumerate}

\subsubsection{Modification to the Convolution Matrix $\matr{H}$}\label{sec:methods:aimx:precorr}

Assuming that the AIM grid has~$N_g$ points, the~$(m,n)$ entry of~$\matr{H}$ is
\begin{align}
  \matr{H}^{(m,n)} = G\left(k_0, \vect{r}_m, \vect{r}_n\right),\label{eq:Hmn}
\end{align}
where ${m,n = 1,2\ldots N_g}$, and $\vect{r}_i$ is the $i^{\mathrm{th}}$ grid point.
Due to the regularity of the AIM grid and the translation-invariance of $\Grrpko$, $\matr{H}$ is a three-level Toeplitz matrix~\cite{pfftmain}.
Only its first row contains unique entries which must be computed and stored.
When ${m=n}$, the source and observation grid points coincide, i.e. ${r = 0}$.
The term $\matr{H}^{(m,m)}$ is not well defined, since ${\Grrpko \rightarrow \infty}$ as ${r \rightarrow 0}$.
However, the ${r = 0}$ case necessarily resides within the near region, and is accounted for with direct integration using appropriate singularity subtraction or cancellation techniques~\cite{gibson}.
Therefore, in the conventional AIM, one can set $\matr{H}^{(m,m)}$ to an arbitrary value, so long as $\matr{H}_{\mathrm{NR}}^{(m,m)}$ is set to the same value.
This ensures that $\matr{H}^{(m,m)}$ is cancelled out when near-region precorrections are applied.

In the proposed AIMx, the $\matr{H}^{(m,m)}$ term must be handled more carefully.
The contribution of $\Grrp[\mathrm{s}]$ in the near region, particularly when ${r=0}$, is accounted for by $\Lmat[\mathrm{s,NR}]$ via direct integration, with the extraction or cancellation of the singularity~\cite{gibson}.
Therefore, $\matr{H}^{(m,m)}_\mathrm{s}$ can be set to an arbitrary value, so long as $\matr{H}^{(m,m)}_\mathrm{s,NR}$ is set to the same value.
For simplicity, we set
\begin{align}
  \matr{H}^{(m,m)}_\mathrm{s} = \matr{H}^{(m,m)}_\mathrm{s,NR} = 0.\label{eq:Hsmm}
\end{align}
The term $\matr{H}^{(m,m)}_\mathrm{d}$ contains the contribution of $\Grrpko[\mathrm{d}]$ for ${r=0}$, which must still be accounted for and can be obtained by inspection of~\eqref{eq:hgftaylor}.
The first term on the right-hand side of~\eqref{eq:hgftaylor} is $\Grrp[\mathrm{s}]$, while the remaining terms represent $\Grrpko[\mathrm{d}]$,
\begin{align}
  \Grrpko[\mathrm{d}] = \frac{\left(-jk_0\right)}{4\pi} + \frac{\left(-jk_0\right)^2}{4\pi\cdot 2!} r + \cdots.\label{eq:hgfdtaylor}
\end{align}
When ${r=0}$, all but the first term on the right-hand side of~\eqref{eq:hgfdtaylor} are $0$.
Therefore, we let
\begin{align}
  \matr{H}^{(m,m)}_\mathrm{d} = G_{\mathrm{d}}\left(k_0, \vect{r}_m, \vect{r}_m\right) = \frac{(-jk_0)}{4\pi}.\label{eq:Hdmm}
\end{align}
Finally, since ${\matr{H}^{(m,m)} = \matr{H}^{(m,m)}_\mathrm{s} + \matr{H}^{(m,m)}_\mathrm{d}}$, using~\eqref{eq:Hsmm} and~\eqref{eq:Hdmm} we have
\begin{align}
  \matr{H}^{(m,m)} = \matr{H}^{(m,m)}_\mathrm{d} = \frac{(-jk_0)}{4\pi},\label{eq:Hmm}
\end{align}
which is the simple modification that must be made to $\matr{H}$ compared to the conventional AIM.
This modification ensures that there is no undefined behavior when ${r = 0}$, and that $\matr{H}$ correctly accounts for frequency-dependent effects for all values of~$r$, including~${r=0}$.

\subsubsection{Preconditioning}\label{sec:methods:aimx:pc}

The fact that only frequency-independent near-region interactions are precomputed with the MoM has important implications for constructing a preconditioner to solve~\eqref{eq:EFIEAIMx} iteratively.
Common choices of preconditioner include diagonal, near-region and block-diagonal entries of $\Lmat(k_0)$~\cite{AIMPC,near_zone_PC}.
The key challenge in preconditioner assembly for the proposed method is that $\Lmat[\mathrm{d}](k_0)$ is not assembled explicitly, and only the entries of $\Lmat[\mathrm{s}]$ are available prior to the iterative solution of~\eqref{eq:EFIEAIMx}.

In the case of diagonal preconditioning, only the self-terms of $\Lmat(k_0)$ are involved, which are dominated by the contribution of $\Grrp[\mathrm{s}]$.
Therefore, it is reasonable to expect that in the proposed method, using the diagonal of $\Lmat[\mathrm{s}]$ as the preconditioner should be just as effective as using the diagonal of $\Lmat(k_0)$, as in the conventional AIM.
This is indeed the case, as demonstrated through several numerical examples in \secref{sec:results}.
The examples show that when a full-wave diagonal preconditioner suffices for a given problem, the proposed preconditioner is equally effective, even at high frequencies.

When considering structures which are beyond tens or hundreds of wavelengths in size, diagonal preconditioners may no longer be sufficient, and more advanced preconditioning or regularization techniques may be required.
For example, in the case of near-region or block-diagonal preconditioning, 
using the entries of $\Lmat[\mathrm{s}]$ in the proposed AIMx may not be as effective as using those of $\Lmat$, as in the conventional AIM.
This is because the off-diagonal entries of $\Lmat(k_0)$ become increasingly influenced by $\Grrpko[\mathrm{d}]$ as $r$ increases.
If off-diagonal entries of $\Lmat(k_0)$ are required for preconditioning, one may use FFTs via~\eqref{eq:LAIMx} to explicitly generate selected entries of $\Lmat[\mathrm{d}](k_0)$ to augment the corresponding entries of $\Lmat[\mathrm{s}]$.
This procedure may be performed at each frequency point, or at every few frequency points.
The associated computational cost is similar to that of the near-region precorrection phase of the conventional AIM, which may be significant but is not typically a bottleneck.
Crucially, the direct integration procedure is still avoided at each new frequency point.
In this work, we demonstrate that a diagonal preconditioner works well for several structures of practical interest.

\subsubsection{Modification for Increased Accuracy}\label{sec:methods:aimx:acc}

\begin{figure}[t]
  \centering
  \includegraphics[width=\linewidth]{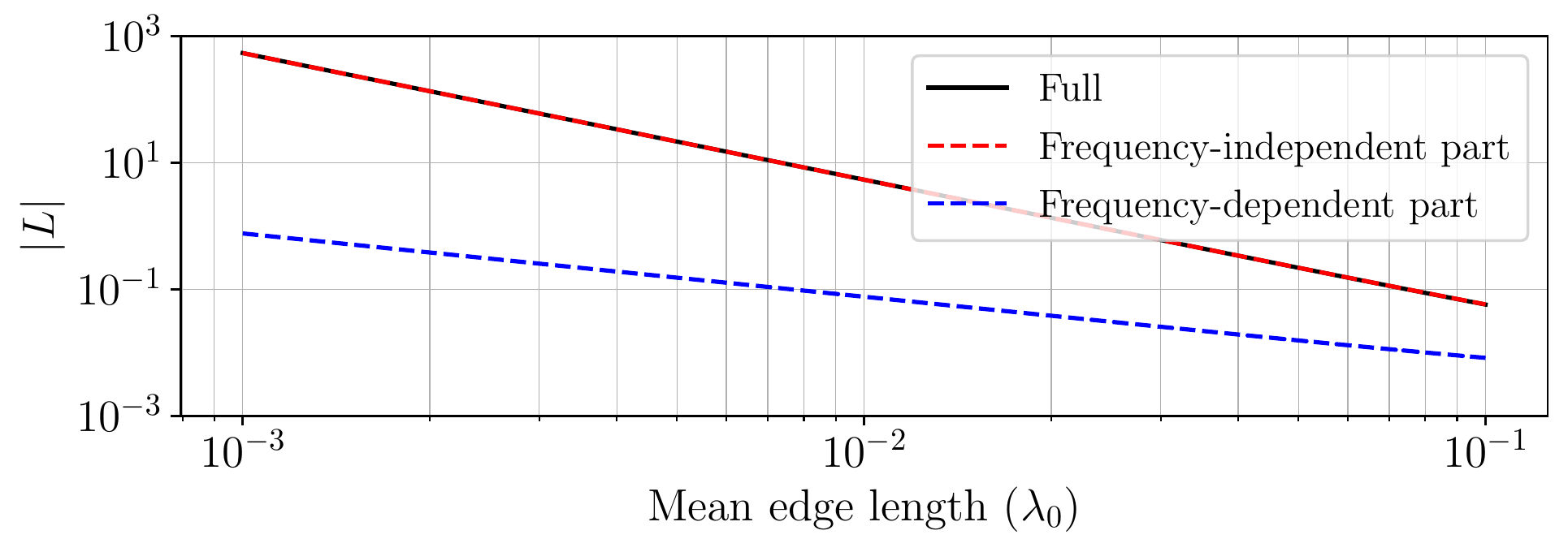}
  \caption{Frequency-dependent and frequency-independent parts of ${\Lmat = \LmatA + k_0^{-2}\LmatPhi}$ as a function of mean edge length.}\label{fig:contributions}
\end{figure}

Even though $\Grrpko[\mathrm{d}]$ is nonsingular, it is non-smooth when ${\vect{r} = \vecprime{r}}$, due to the linear term in~\eqref{eq:hgftaylor},
\begin{align}
  k_0^2\,\Grrp[\mathrm{lin}] = \frac{(-jk_0)^2}{4\pi\cdot 2!}r.\label{eq:linterm}
\end{align}
This is because ${r = \abs{\vect{r}-\vecprime{r}}}$ is nonlinear in the Cartesian coordinates of the test and source points $\vect{r}$ and $\vecprime{r}$.
Therefore, when ${r\to0}$, the terms of~\eqref{eq:hgftaylor} involving odd powers of $r$ may not be captured with very high accuracy by the FFT-based computations on the AIM grid, depending on the number of moments matched\footnote{We are grateful to an anonymous reviewer for pointing this out.}. This occurs when basis and testing functions have an overlap.
However, the numerical results presented in \secref{sec:results} indicate that for many practical engineering applications, the proposed method still achieves excellent overall accuracy.
This is partly because the matrix entries corresponding to overlapping basis and test functions are dominated by the singular term $\Grrp[\mathrm{s}]$, while the contributions of $k_0^2\,\Grrp[\mathrm{lin}]$ and subsequent terms in~\eqref{eq:hgftaylor} tend to $0$ as ${r\to0}$.
As an example, we consider the $\opL$ operator in the EFIE for the PEC sphere described in \secref{sec:validation:self}.
\figref{fig:contributions} shows the magnitude of a single diagonal element (``self term'') of the system matrix ${\Lmat = \LmatA + k_0^{-2}\LmatPhi}$, which corresponds to the case of fully overlapping basis and test functions.
Also plotted are the frequency-dependent and frequency-independent parts of $\Lmat$.
It is clear that the self-term is dominated by the frequency-independent part of $\Lmat$.

When higher accuracy is required, the proposed method can be modified slightly by treating $\Grrp[\mathrm{lin}]$ separately, with direct integration, and redefining $\Grrpko[\mathrm{d}]$ as
\begin{multline}
  \Grrpko[\mathrm{d}] = \Grrpko\\ - \Grrp[\mathrm{s}] - k_0^2\,\Grrp[\mathrm{lin}].
\end{multline}
Accordingly, $\Lmat(k_0)$ in~\eqref{eq:Lmatdecomp} is rewritten as
\begin{align}
  \Lmat(k_0) = \Lmat[\mathrm{s}] + k_0^2\,\Lmat[\mathrm{lin}] + \Lmat[\mathrm{d}](k_0).\label{eq:Lmatdecomplin}
\end{align}
In~\eqref{eq:Lmatdecomplin}, direct integration is required for all near-region entries of $\Lmat[\mathrm{s}]$.
However, for $\Lmat[\mathrm{lin}]$, direct integration is only required only for entries corresponding to overlapping basis and test functions.
The remaining entries of $\Lmat[\mathrm{s}]$ and $\Lmat[\mathrm{lin}]$ are approximated with grid-based FFTs.
The term $\Lmat[\mathrm{lin}]$ is accounted for on-the-fly during the matrix-vector product, scaled by the appropriate value of $k_0^2$.
The effect of this modification on overall accuracy is analyzed numerically in \secref{sec:validation}.
For $N$ mesh edges, the modification described above incurs an $\mathcal{O}(N)$ increase in the CPU time and memory usage.
However, the advantages of the proposed method compared to both the conventional AIM, and the methods in~\cite{ChewAEFIEperturbOrig,ChewAEFIEperturb}, are still maintained.
The accuracy can be further improved by separately treating the quadratic, cubic etc. terms in~\eqref{eq:hgftaylor}, which would lead to another $\mathcal{O}(N)$ increase in memory and CPU time.
However, such a treatment is unnecessary for most practical applications, as demonstrated in \secref{sec:results}.

\subsection{AIMx for the Double-Layer Potential Operator}\label{sec:methods:K}

When~$\mathcal{S}$ is a closed surface, one may also express~$\Jr$ via the magnetic field integral equation (MFIE)~\cite{gibson},
\begin{align}
  -\nhat \times \nhat \times \opK \bigl[\Jrp\bigr]\depk[0] = \nhat \times \nhat \times \Hr[\mathrm{inc}],\label{eq:MFIEout}
\end{align}
where, in free space, the double-layer potential operator~\cite{book:colton} $\opK$ is defined as~\cite{ChewWAF}
\begin{align}
  \opK\bigl[\vect{X}\rp\bigr]\depk[0] = \int_\mathcal{S} \nabla\Grrpko \times \vect{X}\rp\, d\mathcal{S},\label{eq:opK}
\end{align}
where
\begin{align}
  \nabla\Grrpko = -\rhat\,(1 + jk_0r)\frac{e^{-j k_0 r}}{4\pi r^2},\label{eq:ghgf}
\end{align}
and ${\rhat = (\vect{r} - \vect{r}^{\,'})/r}$.
For closed PEC objects, it is often desirable to formulate an integral equation by adding~\eqref{eq:EFIEout} to a scaled version of~\eqref{eq:MFIEout}, to obtain the combined field integral equation (CFIE)~\cite{gibson},
\begin{multline}
  -j\omega\mu_0\, \nhat \times \opL \bigl[\Jrp\bigr]\depk[0]
  - \eta_0\,\nhat \times \nhat \times \opK \bigl[\Jrp\bigr]\depk[0] \\= \nhat \times \Er[\mathrm{inc}] + \eta_0\,\nhat \times \nhat \times \Hr[\mathrm{inc}],\label{eq:CFIEout}
\end{multline}
where~$\eta_0$ is the wave impedance of free space.
Discretizing and testing~\eqref{eq:CFIEout} as before yields its discrete counterpart,
\begin{align}
  \left(-j\omega\mu_0 \Lmat - \eta_0\Kmat \right) \Jmat = \Emat[\mathrm{inc}] + \eta_0\Hmat[\mathrm{inc}],\label{eq:CFIEdis}
\end{align}
where $\Kmat$ is the discretized ${\nhat\times\nhat\times\opK}$ operator, and column vector $\Hmat[\mathrm{inc}]$ contains coefficients of the basis functions associated with $\nhat\times\Hr[\mathrm{inc}]$.

To develop the proposed method for the $\opK$ operator, we expand $\nabla\Grrpko$ in a Taylor series about ${r = 0}$,
\begin{multline}
  \nabla\Grrpko \\= -\frac{\rhat}{4\pi}\left(\frac{1}{r^2} - \frac{\left(-jk_0\right)^2}{2!} - 2\frac{\left(-jk_0\right)^3}{3!} r + \cdots\right).\label{eq:ghgftaylor}
\end{multline}
As in the case of $\Grrpko$, the singularity of $\nabla\Grrpko$ is contained entirely in the first term on the right-hand side of \eqref{eq:ghgftaylor}, which is frequency-independent.
The remaining frequency-dependent terms are either constant or decay to $0$ as ${r \rightarrow 0}$, and are continuous functions for small~$r$.
Following the same approach as in \secref{sec:methods:L} for $\opL$, we express $\opK$ as
\begin{align}
  \opK\bigl[\vect{X}\rp\bigr]\depk[0] = \opK_{\mathrm{s}}\bigl[\vect{X}\rp\bigr] + \opK_{\mathrm{d}}\bigl[\vect{X}\rp\bigr]\depk[0],\label{eq:opKdecomp2}
\end{align}
where the frequency-independent part is
\begin{align}
  \opK_{\mathrm{s}}\bigl[\vect{X}\rp\bigr] = \int_\mathcal{S} \nabla\Grrp[\mathrm{s}] \times \vect{X}\rp\, d\mathcal{S},\label{eq:opKs}
\end{align}
and the frequency-dependent part is
\begin{align}
  \opK_{\mathrm{d}}\bigl[\vect{X}\rp\bigr]\depk[0] = \int_\mathcal{S} \nabla\Grrpko[\mathrm{d}] \times \vect{X}\rp\, d\mathcal{S},\label{eq:opKd}
\end{align}
In \eqref{eq:opKs} and \eqref{eq:opKd},
\begin{align}
  \nabla\Grrp[\mathrm{s}] &= -\frac{\rhat}{4\pi r^2},\label{eq:ghgfs}\\
  \nabla\Grrpko[\mathrm{d}] &= \nabla\Grrpko - \nabla\Grrp[\mathrm{s}].\label{eq:ghgfd}
\end{align}
Therefore, $\Kmat(k_0)$ may be decomposed as
\begin{align}
  \Kmat(k_0) = \Kmat[\mathrm{s}] + \Kmat[\mathrm{d}](k_0),\label{eq:Kmatdecomp}
\end{align}
where $\Kmat[\mathrm{s}]$ and $\Kmat[\mathrm{d}](k_0)$ are the discretized $\opK_{\mathrm{s}}$ and $\opK_{\mathrm{d}}$ operators, respectively.
Following the approach in \secref{sec:methods:L}, $\Kmat(k_0)$ may be approximated as
\begin{align}
  \Kmat(k_0) \approx \Kmat[\mathrm{s,NR}] - \Kmat[\mathrm{s,P}] + \matr{W}^{(\Kmat)}\matr{H}_\nabla(k_0)\matr{P}^{(\Kmat)},\label{eq:KAIMx}
\end{align}
where $\Kmat[\mathrm{s,NR}]$ and $\Kmat[\mathrm{s,P}]$ are the near-region and precorrection matrices, respectively, which are frequency-independent and involve the kernel $\nabla\Grrp[\mathrm{s}]$, and $\matr{H}_\nabla(k_0)$ is the convolution matrix encoding $\nabla\Grrpko$.
Matrices $\matr{W}^{(\Kmat)}$ and $\matr{P}^{(\Kmat)}$ are the interpolation and projection matrices associated with $\Kmat(k_0)$, respectively.
In keeping with the discussion in \secref{sec:methods:aimx:precorr}, the self-term of $\matr{H}_\nabla(k_0)$ should be set as
\begin{align}
  \matr{H}_\nabla^{(0,0)}(k_0) = \nabla G_{\mathrm{d}}\left(k_0, \vect{r}_0, \vect{r}_0\right) = \frac{(-jk_0)^2}{4\pi\cdot2!},\label{eq:gH00}
\end{align}
which is the second term of the expansion~\eqref{eq:ghgftaylor}.
The singularity of $\nabla\Grrp[\mathrm{s}]$ at ${r=0}$ is accounted for with direct integration via $\Kmat[\mathrm{s,NR}]$.

Finally, using \eqref{eq:LAIMx} and \eqref{eq:KAIMx} in \eqref{eq:CFIEdis}, we obtain the AIMx-accelerated CFIE for PECs,
\begin{multline}
  -j\omega\mu_0 \left[\vphantom{\frac{j}{k_0}}\left(\LmatA[\mathrm{s,NR}] - \LmatA[\mathrm{s,P}] + \matr{W}^{(A)} \matr{H} \matr{P}^{(A)}\right) \right. \\ \left. + \frac{1}{k_0^2}\left(\LmatPhi[\mathrm{s,NR}] - \LmatPhi[\mathrm{s,P}] + \matr{W}^{(\phi)} \matr{H} \matr{P}^{(\phi)}\right) \right. \\ \left. - \frac{j}{k_0}\left(\Kmat[\mathrm{s,NR}] - \Kmat[\mathrm{s,P}] + \matr{W}^{(\Kmat)} \matr{H}_\nabla \matr{P}^{(\Kmat)}\right)\right] \Jmat \\= \Emat[\mathrm{inc}] + \eta_0\Hmat[\mathrm{inc}].\label{eq:CFIEAIMx}
\end{multline}
The comments regarding preconditioners in \secref{sec:methods:aimx:pc} apply to \eqref{eq:CFIEAIMx} as well.
It should also be noted that the modification described in \secref{sec:methods:aimx:acc} is not necessary for the double-layer potential operator.
This is because the direct integration is typically performed in a principal value sense, while the self-interactions, including the case when ${\vect{r} = \vecprime{r}}$, are accounted for with an analytical residue term~\cite{ChewWAF}.

\subsection{AIMx for Layered Background Media}\label{sec:methods:aimx:mgf}

For layered background media, we consider the multilayer Green's function (MGF) as described in formulation C of~\cite{MGF02}, though the proposed method may be extended to other formulations as well.
The definition of~$\opL$ becomes~\cite{ChewWAF}
\begin{multline}
  \opL\bigl[\vect{X}\rp\bigr] = \int_\mathcal{S}\Gdyadicrrpko[A]\cdot\vect{X}\rp\,d\mathcal{S} \\+ \frac{1}{k_0^2}\int_\mathcal{S}\nabla'\nabla' \cdot \vect{X}\rp\Grrpko[\phi]\,d\mathcal{S},\label{eq:opLmgf}
\end{multline}
where $\Gdyadicrrpko[A]$ and $\Grrpko[\phi]$ are the dyadic and scalar potential Green's functions of the background medium, respectively.
The term $\Grrpko[\phi]$ and the non-zero components of $\Gdyadicrrpko[A]$ are expressed in terms of semi-infinite integrals in the complex plane, which are known as ``Sommerfeld integrals''~\cite{MGF01,SI_PE}.
These integrals are expensive to compute, even when approximations such as the discrete complex image method (DCIM)~\cite{DCIM01} are employed.
As a result, the near-region computations involving direct integration are particularly time consuming for layered media.

In the following, $\Grrpkom$ represents either $\Grrpko[\phi]$, or a non-zero component of $\Gdyadicrrpko[A]$.
Regardless of the method used to compute $\Grrpkom$, it is common practice to extract the quasistatic contributions in the spectral domain, which are then added back analytically in spatial domain~\cite{qse,SI_PE,DCIM01}.
This ensures that the integrands involved in the computation of $\Grrpkom$ decay sufficiently fast~\cite{qse}.
Extracting quasistatic terms also provides a simple way to compute the frequency-independent part of $\Grrpkom$, as discussed below.
Therefore, we may write
\begin{align}
  \Grrpkom = \Grrpkom[\mathrm{q}] + \Grrpkom[\mathrm{r}],\label{eq:qse}
\end{align}
where $\Grrpkom[\mathrm{q}]$ is the quasistatic part of $\Grrpkom$, and $\Grrpkom[\mathrm{r}]$ is the remainder.
In general, $\Grrpkom[\mathrm{q}]$ may be expressed as~\cite{qse}
\begin{align}
  \Grrpkom[\mathrm{q}] = \sum_{n=1}^{N_s} \Grrpm[\mathrm{s}n] + \sum_{n=1}^{N_q} \Grrpkom[\mathrm{q}n],\label{eq:qmgfdef}
\end{align}
where $N_s$, $N_q$ and the forms of $\Grrpm[\mathrm{s}n]$ and $\Grrpkom[\mathrm{q}n]$ all depend on the location of source and observation points~\cite{qse}.
In~\eqref{eq:qmgfdef}, $\Grrpkom[\mathrm{q}]$ contains some terms, $\Grrpm[\mathrm{s}n]$, that are already frequency-independent, such as the quasistatic parts of the off-diagonal components of \Gdyadicrrpko[A].
The other terms, $\Grrpkom[\mathrm{q}n]$, depend on frequency, and have the form~\cite{qse},
\begin{align}
  \Grrpkom[\mathrm{q}n] = \frac{C_n}{4\pi} \frac{e^{-jk_i\alpha_n}}{\alpha_n},\label{eq:qs1}
\end{align}
where $C_n$ is a constant which depends on the material parameters of the layered medium~\cite{qse}, $k_i$ is the wave number in the layer in which the source resides, ${\rho = \sqrt{(x-x')^2 + (y-y')^2}}$, and $\alpha_n = \sqrt{\rho^2 + \gamma_n^2}$.
Term $\gamma_n$ is a linear function of $z$, $z'$ and the elevations and heights of source and observation layers~\cite{qse}.
Equation \eqref{eq:qs1} has a form similar to \eqref{eq:hgf}, and the Taylor expansion of $\Grrpkom[\mathrm{q}n]$ about ${\alpha = 0}$ reads
\begin{multline}
  \Grrpkom[\mathrm{q}n] = C_n \left[ \frac{1}{4\pi\alpha_n} + \frac{\left(-jk_i\right)}{4\pi} \right. \\[1mm] \left. + \frac{\left(-jk_i\right)^2}{4\pi\cdot2!} \alpha_n + \frac{\left(-jk_i\right)^3}{4\pi\cdot3!} \alpha_n^2 + \cdots\right],\label{eq:mgftaylor}
\end{multline}
where again the singularity is contained in the first term, which is frequency-independent.
We may decompose $\Grrpkom[\mathrm{q}n]$ into frequency-independent and frequency-dependent parts as
\begin{align}
  \Grrpkom[\mathrm{q}n] = \underbrace{\vphantom{\left(\frac{1}{\alpha}\right)}\frac{C_n}{4\pi\alpha_n}}_{\mathclap{\Grrpm[\mathrm{qs}n]}} + \underbrace{\frac{C_n}{4\pi}\left(\frac{e^{-jk_i\alpha_n}}{\alpha_n} - \frac{1}{\alpha_n}\right)}_{\mathclap{\Grrpkom[\mathrm{qd}n]}}.\label{eq:qmgf1decomp}
\end{align}
Using~\eqref{eq:qmgf1decomp} and~\eqref{eq:qmgfdef} in~\eqref{eq:qse}, we get
\begin{multline}
  \Grrpkom = \sum_{n=1}^{N_s} \Grrpm[\mathrm{s}n] + \sum_{n=1}^{N_q} \Grrpm[\mathrm{qs}n] \\+ \sum_{n=1}^{N_q} \Grrpkom[\mathrm{qd}n] + \Grrpkom[\mathrm{r}],\label{eq:qmgfdef2}
\end{multline}
where the first two terms on the right-hand side are frequency-independent and can be denoted as $\Grrpm[\mathrm{s}]$, while the last two terms are frequency-dependent and are denoted as $\Grrpkom[\mathrm{d}]$.
Therefore, \eqref{eq:qmgfdef2} may be simplified as
\begin{align}
  \Grrpkom = \Grrpm[\mathrm{s}] + \Grrpkom[\mathrm{d}],\label{eq:qse1}
\end{align}
which resembles the free space case in~\eqref{eq:hgfdecomp0}.
This expression holds for both $\Grrpko[\phi]$ and any non-zero component of $\Gdyadicrrpko[A]$.
Equation~\eqref{eq:qse1} indicates that the main ideas of the proposed method also apply to layered media:
\begin{itemize}
  \item the singular behavior of $\Grrpkom$ is entirely contained in the frequency-independent term $\Grrpm[\mathrm{s}]$;
  \item the frequency-dependent term $\Grrpkom[\mathrm{d}]$ is continuous, and decays to~$0$ as ${r \rightarrow 0}$~\cite{qse,SI_PE}, therefore can be approximated by interpolation even in the near region.
\end{itemize}

When the quasistatic part of the MGF is expressed as per formulation C in~\cite{MGF02}, $C_n$ may depend on the complex permittivity of the associated layer, which is frequency-dependent if the layer is lossy.
In this case, $\Grrpm[\mathrm{qs}n]$ may no longer be frequency-independent.
A complete treatment of the lossy layer case is beyond the scope of this paper, and will be the subject of future work.
Here, we assume that the layers are either lossless, or that their losses are small enough that the imaginary part of the complex permittivity can be neglected when~\eqref{eq:qs1} is evaluated in the near region.
As shown in the numerical example in \secref{sec:results:ind}, this approximation still yields accurate results for structures of practical importance.

When the source and observation points both lie at a layer interface, additional singularities arise in the off-diagonal components of $\Gdyadicrrpko[A]$.
This occurs when either the basis or test function has a non-zero~$z$ component, where~$z$ is the direction of stratification.
Here, we consider cases where objects may lie in any layer, but cannot traverse a layer interface.
If an object traverses the interface between two layers, it is divided into two separate objects, each one contained in a single layer.
This ensures that no mesh triangle straddles a layer interface.
Therefore, any testing or basis functions which are exactly at a layer interface can only have~$x$ or~$y$ components, and the additional singularities in the off-diagonal dyadic components do not arise.
However, when the quadrature points are near (but not on) an interface, the off-diagonal dyadic terms may be nearly singular, and we do not extract this near-singularity here.
We have found that the accuracy of the proposed method is not significantly affected by this, as shown in the results in \secref{sec:results:dip} and \secref{sec:results:ind}.

The AIMx concept may also be extended to the curl of $\Gdyadicrrpko[A]$, to handle the $\opK$ operator for layered media~\cite{ChewWAF}.
The proposed AIMx-based EFIE,~\eqref{eq:EFIEAIMx}, and CFIE,~\eqref{eq:CFIEAIMx}, can therefore be applied for PEC objects in layered media.
An important advantage of the proposed method for layered media is that the near-region matrix entries are extremely simple to compute, by taking the static limit of $\Grrpkom[\mathrm{q}]$.
These entries do not require computing Sommerfeld integrals, nor do they require applying approximations such as the DCIM.
Therefore, the near-region computations in the proposed method are significantly more efficient, robust, and easy to implement, compared to the conventional AIM.

\section{Error Analysis}\label{sec:validation}

The proposed method rests on the hypothesis that the frequency-dependent parts of $\opL$ and $\opK$ can be approximated with moment matching on the AIM grid in both the near and far regions.
In this section, this hypothesis is validated for each type of kernel considered through numerical error analysis.
The moment-matching procedure of the AIM~\cite{AIMbles} is equivalent to polynomial interpolation of the kernel, if the basis and test functions have no overlap~\cite{pfftcompare}.
Therefore, when the test function is in the near-region of the basis function but does not overlap with it, the accuracy of the proposed method can be analyzed by considering polynomial interpolation of the kernel.
When the basis and test functions have an overlap, the equivalence to polynomial interpolation is no longer valid\footnote{We are grateful to an anonymous reviewer for pointing this out.}.
In this case, we directly study the errors in the matrix elements computed for a numerical example.
Finally, we report the error in computing the radar cross section of a sphere, compared to the conventional AIM and to the Mie series.
The case of layered media is based on the Taylor expansion of a term whose form resembles the free space Green's function (see~\eqref{eq:qs1}).
Therefore, although the proposed error analysis focuses on the free space case, it is indicative of the error behaviour even in the case of layered media.

\subsection{Non-overlapping basis and test functions in the near-region}\label{sec:validation:nonself}

In this case, we use Lagrange polynomial interpolation of $\Grrpko[\mathrm{d}]$ and $\nabla\Grrpko[\mathrm{d}]$ to demonstrate that both $\opL_{\mathrm{d}}$ and $\opK_{\mathrm{d}}$, respectively, can be approximated accurately with grid-based moment-matching in the near-region, for non-overlapping mesh elements.
For simplicity, we assume that ${\vect{r}\,' = 0}$, and ${r = \abs{\vect{r}} > 0}$.
In addition, we normalize any physical length, say $r$, as ${\theta_{r} \triangleq k_0r}$.

In the following, $\Ktheta{r}$ can represent either $\Gtheta[\mathrm{d}]{r}$ or $\nabla\Gtheta[\mathrm{d}]{r}$.
Consider a regular AIM grid with a stencil which has ${n+1}$ points in each direction.
Given ${n+1}$ samples,
\begin{align}
	(r_0, \Ktheta{r_0}), (r_1, \Ktheta{r_1}) \ldots (r_n, \Ktheta{r_n}),
\end{align}
the Lagrange polynomial approximation of $\Ktheta{r}$ can be written as~\cite{abrstegun}
\begin{align}
  \tKtheta[n]{r} = \sum_{q=0}^n \green{\Ktheta{r_q}} \left( \prod_{0 \leq m \leq n}^{m \neq q} \frac{r - r_m}{r_q - r_m} \right).
\end{align}
The order~$n$ interpolation error is
\begin{align}
  \Delta \Ktheta[n]{r} = \abs{\Ktheta{r} - \tKtheta[n]{r}}.
\end{align}
The goal is to show that ${\Delta \Ktheta[n]{r}}$ is bounded in the near-region for non-overlapping mesh elements, and is controllable via grid refinement.
Following the approach in~\cite{TAPAIMin}, ${\Delta \Ktheta[n]{r}}$ can be bound as~\cite{abrstegun}
\begin{align}
  \frac{\Delta \Ktheta[n]{r}}{k_{0}} \leq \frac{{\left(\theta_{r_n} - \theta_{r_0}\right)}^{\mathrlap{n+1}}}{\left(n+1\right)!} \max_{\theta_{r_0} \leq \theta_{\xi} \leq \theta_{r_n}} \abs{ \frac{d^{(n+1)}\Ktheta{\xi}}{d\theta_\xi^{(n+1)}}},\label{eq:Lbounds}
\end{align}
where~$\theta_{r_0}$ and~$\theta_{r_n}$ are the stencil coordinates nearest to and farthest from the source point, respectively.
The right-hand side of~\eqref{eq:Lbounds} is the error bound, $B_n(\theta_r)$.
When~${\Ktheta{r} = \Gtheta[\mathrm{d}]{r}}$,
\begin{multline}
	\frac{d^{(n+1)}\Ktheta{\xi}}{d\theta_\xi^{(n+1)}} = \frac{(-1)^{n+1}}{4\pi\theta_\xi^{n+2}}\,(n+1)!\,\cdot\\
	\left[ -1 + \sum_{l=0}^{n+1} \begin{pmatrix}n+1 \\ l\end{pmatrix} \frac{(n+1-l)!}{(n+1)!}\, (-j\theta_\xi)^l\, e^{-j\theta_\xi} \right],\label{eq:Dhgf}
\end{multline}
and when~${\Ktheta{r} = \nabla\Gtheta[\mathrm{d}]{r}}$,
\begin{multline}
	\frac{d^{(n+1)}\Ktheta{\xi}}{d\theta_\xi^{(n+1)}} = \frac{-\hat{\theta}}{4\pi}\,
	\sum_{l=0}^{n+1} \begin{pmatrix}n+1 \\ l\end{pmatrix} \frac{(-n-2+l)_{n+1-l}}{\theta_\xi^{n+3-l}}\,\cdot\\ \left[ -\delta_l + (-j)^l\, e^{-j\theta_\xi} \left(1 + j\theta_\xi - l\right) \right].\label{eq:Dghgf}
\end{multline}
In~\eqref{eq:Dghgf}, ${\hat{\theta} = k_0\rhat}$, the notation $(a)_{b}$ indicates the Pochhammer function~\cite{abrstegun}, and ${\delta_l = 1}$ if ${l=0}$, and is $0$ otherwise.

\begin{figure}[t]
	\centering
	\includegraphics[width=\linewidth]{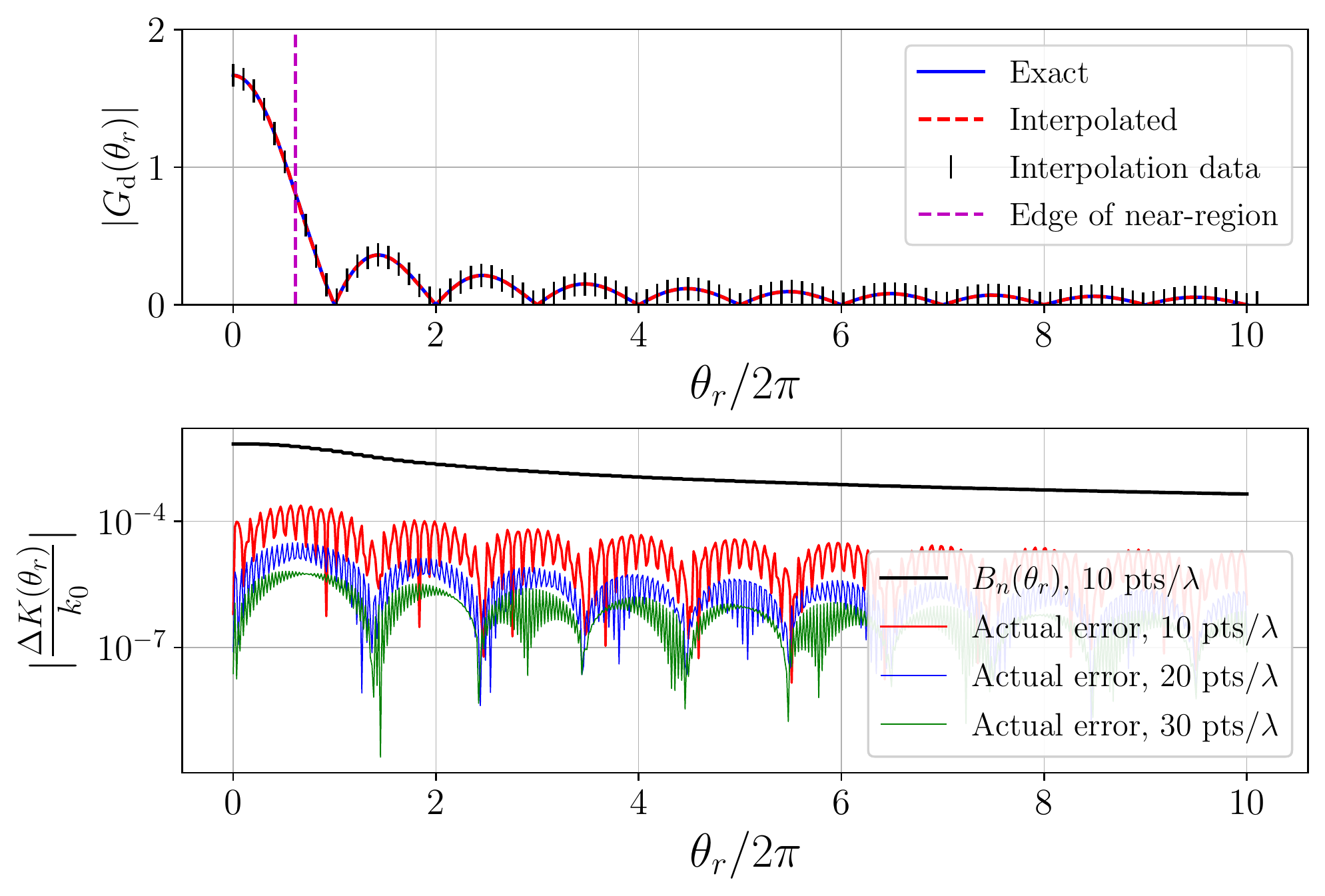}
	\caption{Top panel: interpolation of $\Gtheta[\mathrm{d}]{r}$ for ${n=2}$. Bottom panel: numerical error, and $B_n(\theta_r)$ computed via~\eqref{eq:Dhgf}.}\label{fig:dhgferr}
\end{figure}

\begin{figure}[t]
	\centering
	\includegraphics[width=\linewidth]{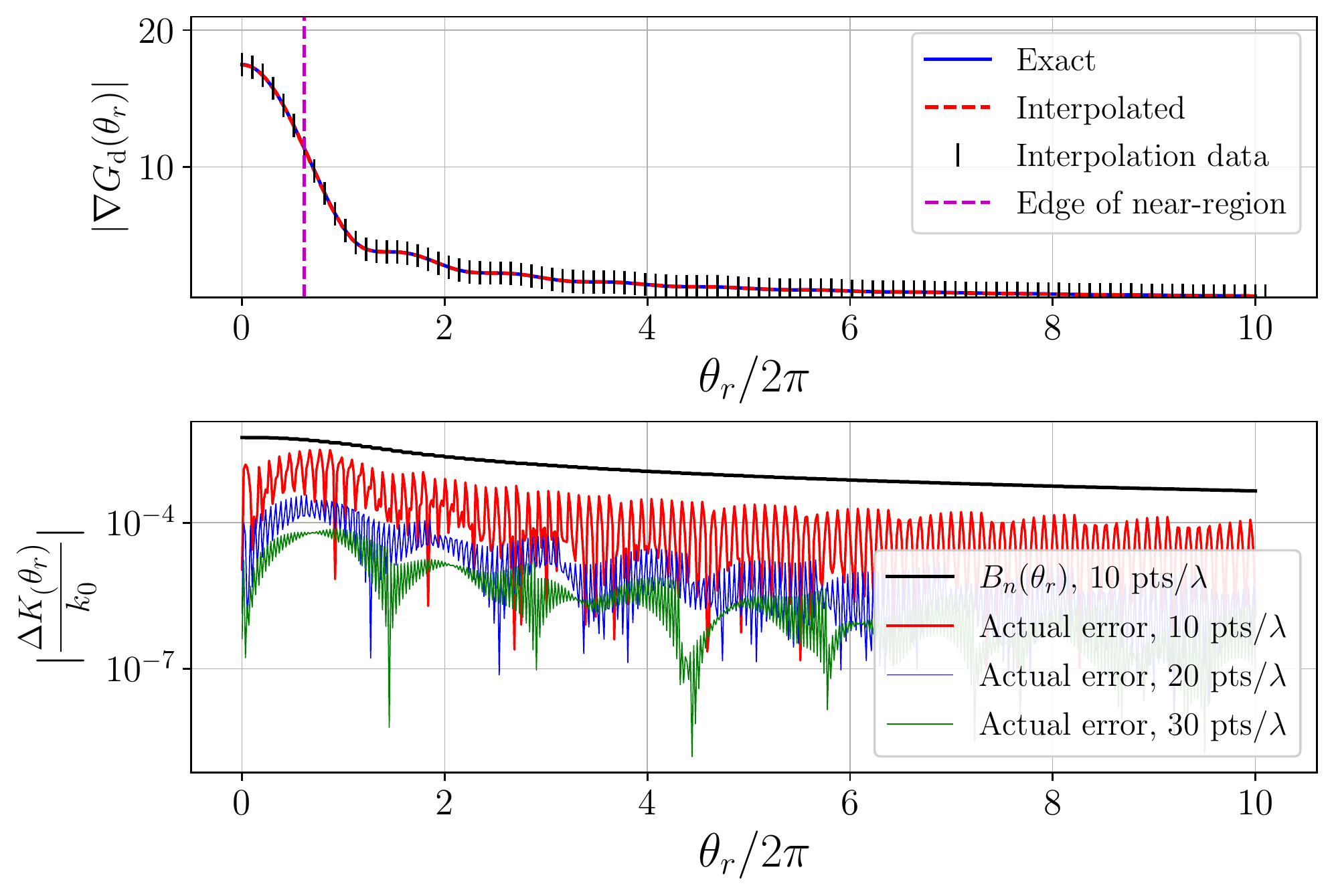}
	\caption{Top panel: interpolation of $\nabla\Gtheta[\mathrm{d}]{r}$ for ${n=2}$. Bottom panel: numerical error, and $B_n(\theta_r)$ computed via~\eqref{eq:Dghgf}.}\label{fig:dghgferr}
\end{figure}

\figref{fig:dhgferr} shows, over a wide range of electrical sizes, the results of the error analysis for $\Gtheta[\mathrm{d}]{r}$, for ${n=2}$.
\figref{fig:dghgferr} shows the same for $\nabla\Gtheta[\mathrm{d}]{r}$.
In both cases, the top panel shows the exact and interpolated curves, for the given sample points indicated with vertical black bars.
It is assumed that $10$ samples are used per wavelength.
The magenta dashed line shows a typical choice for the size of the near region.
The bottom panel shows the computed numerical error, as well as the analytical bound~$B_n(\theta_r)$.
\figref{fig:dhgferr} and \figref{fig:dghgferr} clearly indicate that the error is bounded as~${\theta_r \rightarrow 0}$, and is not significantly larger within the near region, as compared to the region just beyond.
Also plotted in the bottom panel is the computed numerical error when the number of samples is increased to $20$ and $30$ points per wavelength, confirming that the error is controllable via grid refinement.
Therefore, for non-overlapping basis and testing functions in the near-region, the frequency-dependent part of the free space Green's function, and its gradient, can be interpolated with polynomials.

\subsection{Overlapping basis and test functions in the near-region}\label{sec:validation:self}

To study the errors when basis and test functions overlap, we consider the EFIE~\eqref{eq:EFIEdis} for a PEC sphere with diameter~$1\,$m, in free space.
The sphere is meshed with~$3{,}080$ triangles and~$4{,}620$ edges, and an AIM grid with ${25 \times 25 \times 25}$ points along~$x$,~$y$ and~$z$, respectively, was used with~${n=2}$.
The maximum relative error in the diagonal matrix elements of $\LmatA$ and $\LmatPhi$, which correspond to a full overlap in basis and test functions, is computed as
\begin{align}
	\Delta L = \max\,\mathrm{diag}\,\frac{\abs{\Lmat_{\mathrm{AIMx}} - \Lmat_{\mathrm{direct}}}}{\abs{\Lmat_{\mathrm{direct}}}},
\end{align}
where $\Lmat$ represents either $\LmatA$ or $\LmatPhi$. Subscripts ``AIMx'' and ``direct'' indicate that matrix entries were computed via the proposed method or via direct integration, respectively.
We also report $\Delta L$ for the full EFIE system matrix ${\LmatA + k_0^{-1}\LmatPhi}$.
Typically, structures are meshed so that the mean edge length is no larger than approximately $10^{-1}\lambda_0$, where $\lambda_0$ is the wavelength in free space.
Therefore, $\Delta L$ is computed for frequencies between $5\,$MHz and $500\,$MHz, so that the mean edge length ranges from $10^{-3}\lambda_0$ to $10^{-1}\lambda_0$.

\begin{figure}[t]
	\centering
	\includegraphics[width=\linewidth]{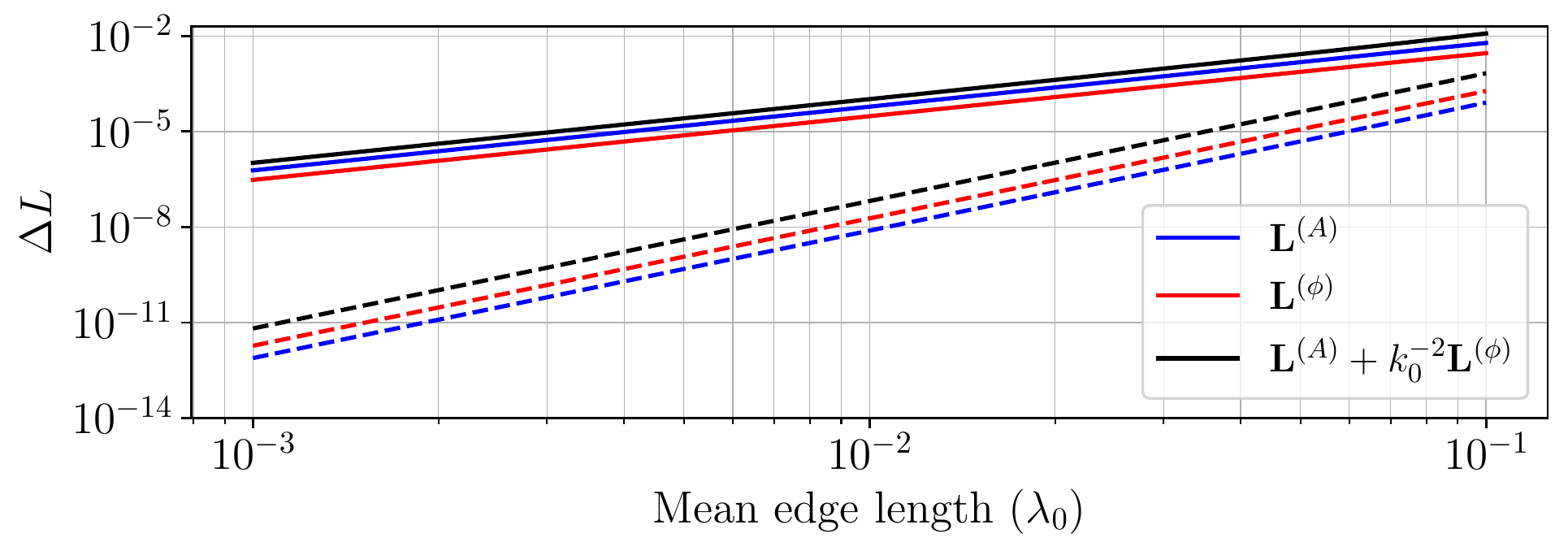}
	\caption{Relative error for overlapping basis and test functions with the proposed method (solid lines) and the proposed method with the modification in \secref{sec:methods:aimx:acc} (dashed lines).}\label{fig:selferrors}
\end{figure}

\figref{fig:selferrors} shows $\Delta L$ for $\LmatA$, $\LmatPhi$, and ${\LmatA + k_0^{-1}\LmatPhi}$.
The solid lines correspond to the proposed method, while the dashed lines denote the proposed method with the linear term $\Grrp[\mathrm{lin}]$ treated separately, as suggested in \secref{sec:methods:aimx:acc}.
Even with a relatively coarse mesh with a mean edge length of $10^{-1}\lambda_0$, the error in the proposed method remains below $1\%$, and can be reduced below $10^{-5}$.
For many practical engineering applications, this level of accuracy in the computation of the $\opL$ operator is more than adequate, as shown in the numerical examples in \secref{sec:results}.
\figref{fig:selferrors} also suggests that the modification in \secref{sec:methods:aimx:acc} can be used as a means of error control to achieve an improvement in accuracy, when needed.
It should be noted that this modification was not used in any of the numerical examples considered in \secref{sec:results}.

\subsection{Overall error for a numerical example}\label{sec:validation:sph}

\begin{figure}[t]
  \centering
  \includegraphics[width=\linewidth]{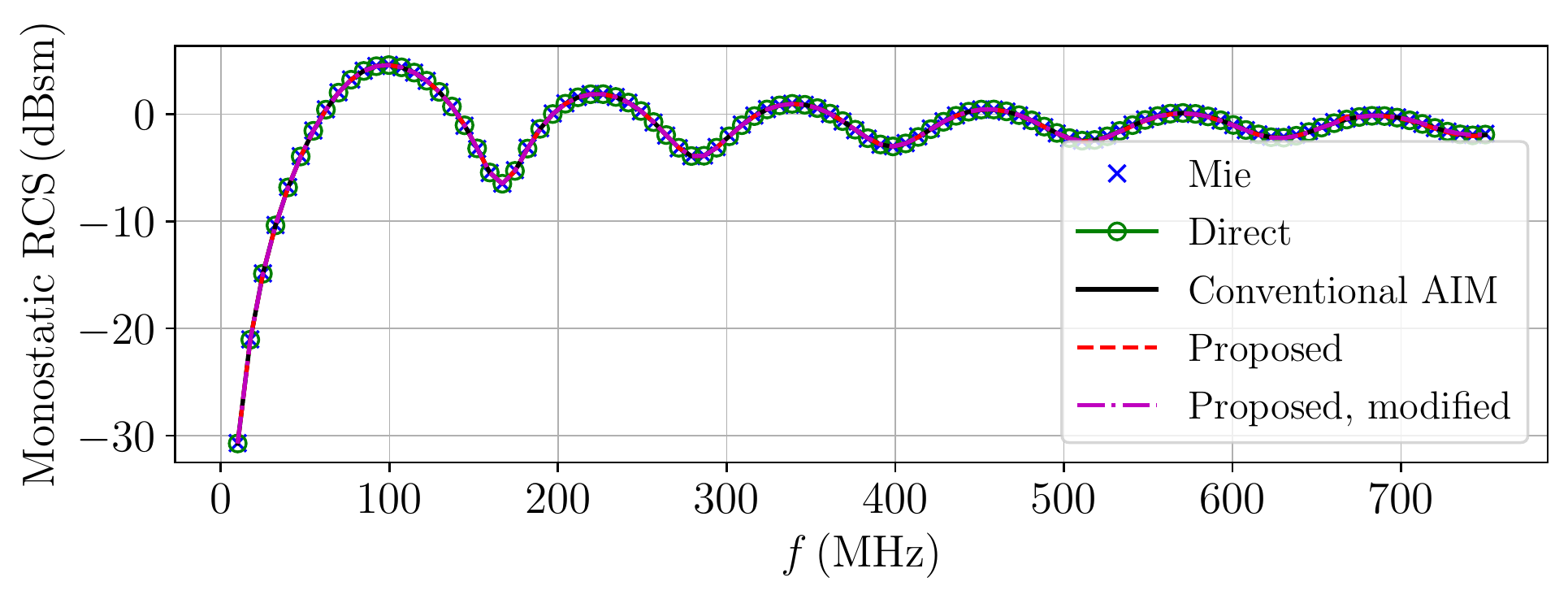}\label{fig:sph:mrcs}\\
  \includegraphics[width=\linewidth]{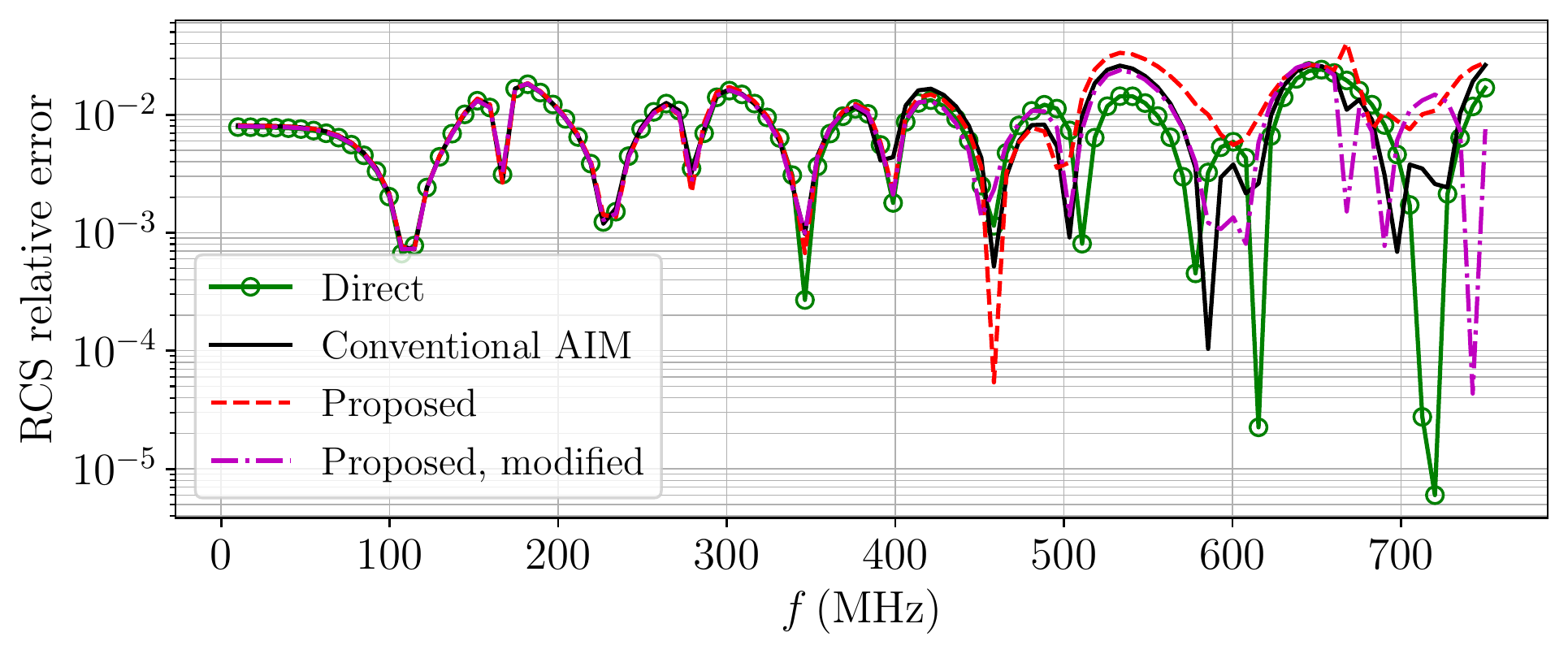}
  \caption{Accuracy validation for the sphere in \secref{sec:validation:sph}. Top panel: monostatic RCS. Bottom panel: relative error.}\label{fig:sph:rcs}\vspace{4mm}
\end{figure}

We next allow a plane wave, with the electric field polarized along the $y$ axis, to impinge on the sphere described above in \secref{sec:validation:self}.
The EFIE system~\eqref{eq:EFIEAIMx} was solved with a diagonal preconditioner.
The top panel of \figref{fig:sph:rcs} shows the monostatic radar cross section (RCS) for $100$ frequency points spaced approximately $17.5\,$MHz apart, computed via the Mie series, direct integration with factorization, the conventional AIM, the proposed method, and the proposed method with the modification in \secref{sec:methods:aimx:acc}.
Excellent agreement between all methods is observed across the entire frequency range.
The bottom panel of \figref{fig:sph:rcs} shows the relative error in the monostatic RCS for each method.
For low and middle frequencies, the error associated with the proposed method is not larger than those of direct integration and the conventional AIM.
At high frequencies, the error in the proposed method is still comparable to that of the other methods.
Furthermore, treating $\Grrp[\mathrm{lin}]$ as described in \secref{sec:methods:aimx:acc} results in an improvement in accuracy at higher frequencies, as expected.
This numerical example illustrates that the proposed method yields an accuracy comparable to that of existing methods, as long as the electrical size is not extremely large.
The results in \secref{sec:results} indicate that the proposed method remains sufficiently accurate for electrical sizes of up to $7\lambda_0$--$10\lambda_0$.

\begin{table}[t]
  \captionsetup{width=0.9\linewidth}
  \centering
  \caption{Dielectric layer configurations for the numerical examples in~\secref{sec:results}. Layers are non-magnetic.}
  \begin{tabular}{cc|ccc}
    \toprule
    \multicolumn{2}{c}{Dipole Array (\secref{sec:results:dip})} & \multicolumn{3}{c}{Inductor Array (\secref{sec:results:ind})} \\
    \cmidrule(lr){1-2}\cmidrule(lr){3-5}
    $\epsilon_r$ & height\,(mm) & $\epsilon_r$ & $\sigma\,$(S/m) & height\,($\mu$m) \\
    \midrule
    \multicolumn{1}{c}{Air} & \multicolumn{1}{c|}{$\infty$} & \multicolumn{2}{c}{Air} & \multicolumn{1}{c}{$\infty$}\\
    \midrule
    & & $2.1$ & $0$ & $6$  \\
    & & $3.7$ & $0$ & $11$ \\
    $2.17$ & $0.254$ & $2.1$ & $0$ & $12$ \\
    $1.04$ & $5.77$ & $3.7$ & $0$ & $11$ \\
    & & $9.0$ & $0$ & $25$ \\
    & & $11.9$ & $10$ & $20$ \\
    \midrule
    \multicolumn{1}{c}{PEC} & \multicolumn{1}{c|}{$\infty$} & \multicolumn{2}{c}{PEC} & \multicolumn{1}{c}{$\infty$}\\
    \bottomrule
  \end{tabular}
  \label{tab:layers}
\end{table}

\section{Results}\label{sec:results}

\begin{table*}[t]
  \centering
  \caption{Performance comparison for the numerical examples in \secref{sec:results}.}
  \begin{tabular}{l|cc|cc|cc|cc|cc}
    \toprule
    \multicolumn{1}{l}{*averaged over all} & \multicolumn{2}{c}{Sphere} & \multicolumn{2}{c}{NASA Almond} & \multicolumn{2}{c}{Reflectarray} & \multicolumn{2}{c}{Dipole Array} & \multicolumn{2}{c}{Inductor Array} \\
    \multicolumn{1}{l}{\ frequency points} & \multicolumn{2}{c}{(\secref{sec:results:sph})} & \multicolumn{2}{c}{(\secref{sec:results:alm})} & \multicolumn{2}{c}{(\secref{sec:results:ref})} & \multicolumn{2}{c}{(\secref{sec:results:dip})} & \multicolumn{2}{c}{(\secref{sec:results:ind})} \\
    \cmidrule(lr){2-3}\cmidrule(lr){4-5}\cmidrule(lr){6-7}\cmidrule(lr){8-9}\cmidrule(lr){10-11}
    & AIM & AIMx & AIM & AIMx & AIM & AIMx & AIM & AIMx & AIM & AIMx \\
    \midrule
    \multicolumn{1}{l|}{Number of mesh elements} & \multicolumn{2}{c|}{$3{,}080$} & \multicolumn{2}{c|}{$40{,}194$} & \multicolumn{2}{c|}{$254{,}736$} & \multicolumn{2}{c|}{$25{,}042$} & \multicolumn{2}{c}{$42{,}396$} \\
    One-time static matrix-fill & -\,- & $2.7\,$sec & -\,- & $1.4\,$min & -\,- & $0.5\,$hr & -\,- & $2.2\,$min & -\,- & $0.2\,$hr \\
    Per-frequency matrix-fill* & $3.9\,$sec & -\,- & $1.4\,$min & -\,- & $1.4\,$hr & -\,- & $3.5\,$min & -\,- & $0.3\,$hr & -\,- \\
    Total CPU time & $34.5\,$min & $11.8\,$min & $515.6\,$min & $117.5\,$min & $120.1\,$hr & $7.5\,$hr & $23.5\,$hr & $7.1\,$hr & $24.2\,$hr & $3.5\,$hr \\
    \multicolumn{1}{l|}{Overall speed-up} & \multicolumn{2}{c|}{$2.9\times$} & \multicolumn{2}{c|}{$4.4\times$} & \multicolumn{2}{c|}{$16.1\times$} & \multicolumn{2}{c|}{$3.3\times$} & \multicolumn{2}{c}{$6.9\times$} \\
    \bottomrule
  \end{tabular}
  \label{tab:prof}
\end{table*}

The accuracy and efficiency of the proposed method are demonstrated in this section, through comparisons with the conventional AIM.
All simulations were performed single-threaded on a 3\,GHz Intel Xeon CPU.
We used PETSc~\cite{petsc-web-page} for sparse matrix manipulation.
The GMRES iterative solver~\cite{gmres} available in PETSc was used to solve~\eqref{eq:EFIEAIMx} and~\eqref{eq:CFIEAIMx}.
A relative residual norm of $10^{-4}$ was used as the convergence tolerance.

\subsection{Sphere}\label{sec:results:sph}

\begin{figure}[t]
  \centering
  \includegraphics[width=\linewidth]{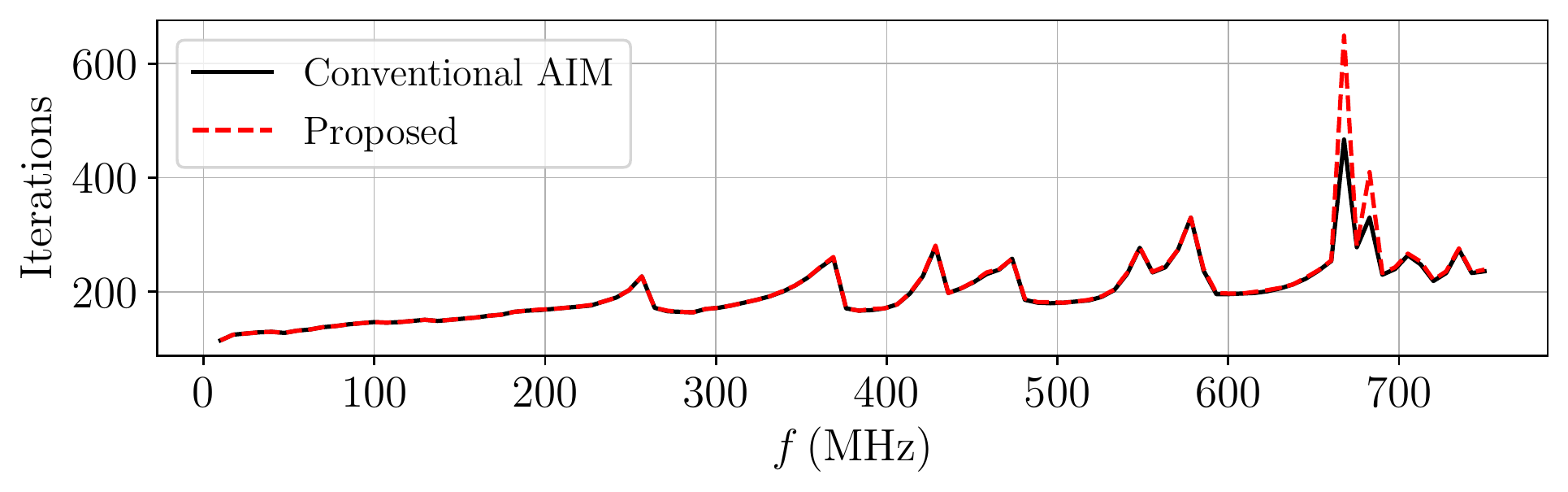}\\
  \includegraphics[width=\linewidth]{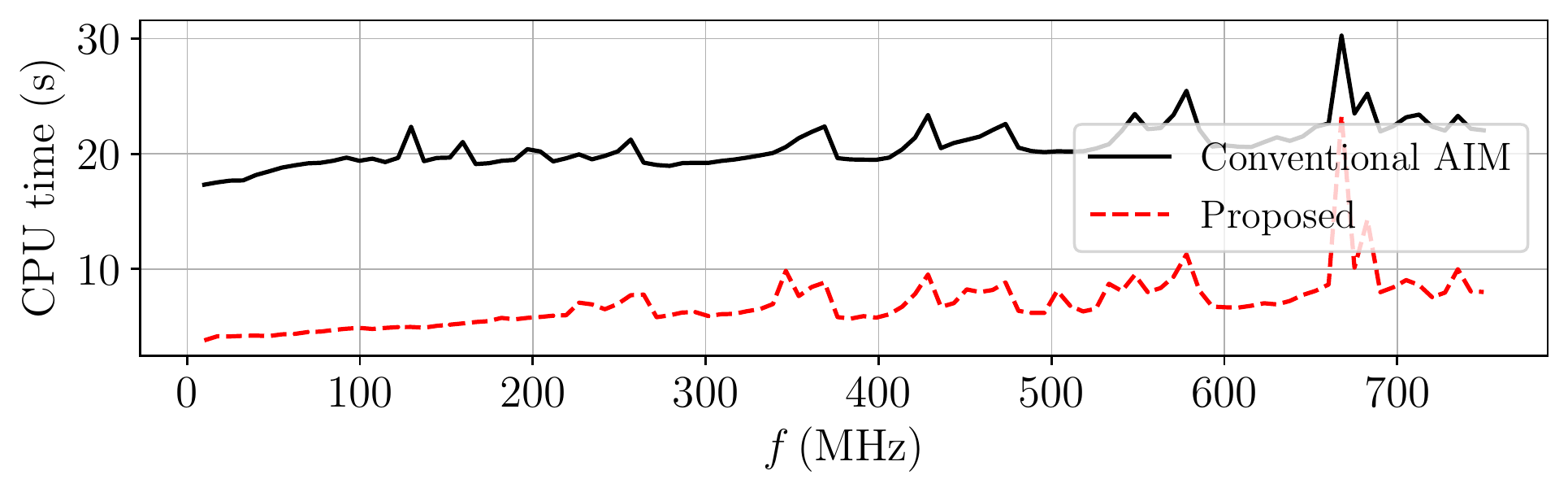}
  \caption{Performance comparison for the sphere in \secref{sec:validation:self}. Top panel: iterations required for convergence. Bottom panel: CPU time per frequency.}\label{fig:sph:prof}
\end{figure}

In this section, we report the performance of the proposed method for the sphere described in \secref{sec:validation:self} and \secref{sec:validation:sph}.
The top panel of \figref{fig:sph:prof} shows that the proposed method requires the same number of iterations as the conventional AIM except at two frequency points, which are near internal resonances.
Therefore, the frequency-independent diagonal preconditioner discussed in \secref{sec:methods:aimx:pc} is nearly as effective as the full-wave diagonal preconditioner across the entire frequency range.
The bottom panel of \figref{fig:sph:prof} shows the CPU time per frequency, and demonstrates the time saved by avoiding the near-region direct integration and precorrection steps at each frequency.
The time taken for near-region computations in both AIMx and the conventional AIM is reported in \tabref{tab:prof}, and a total speed-up of $2.9\times$ was observed.

\subsection{NASA Almond}\label{sec:results:alm}

\begin{figure}[t]
  \centering
  \includegraphics[width=.9\linewidth]{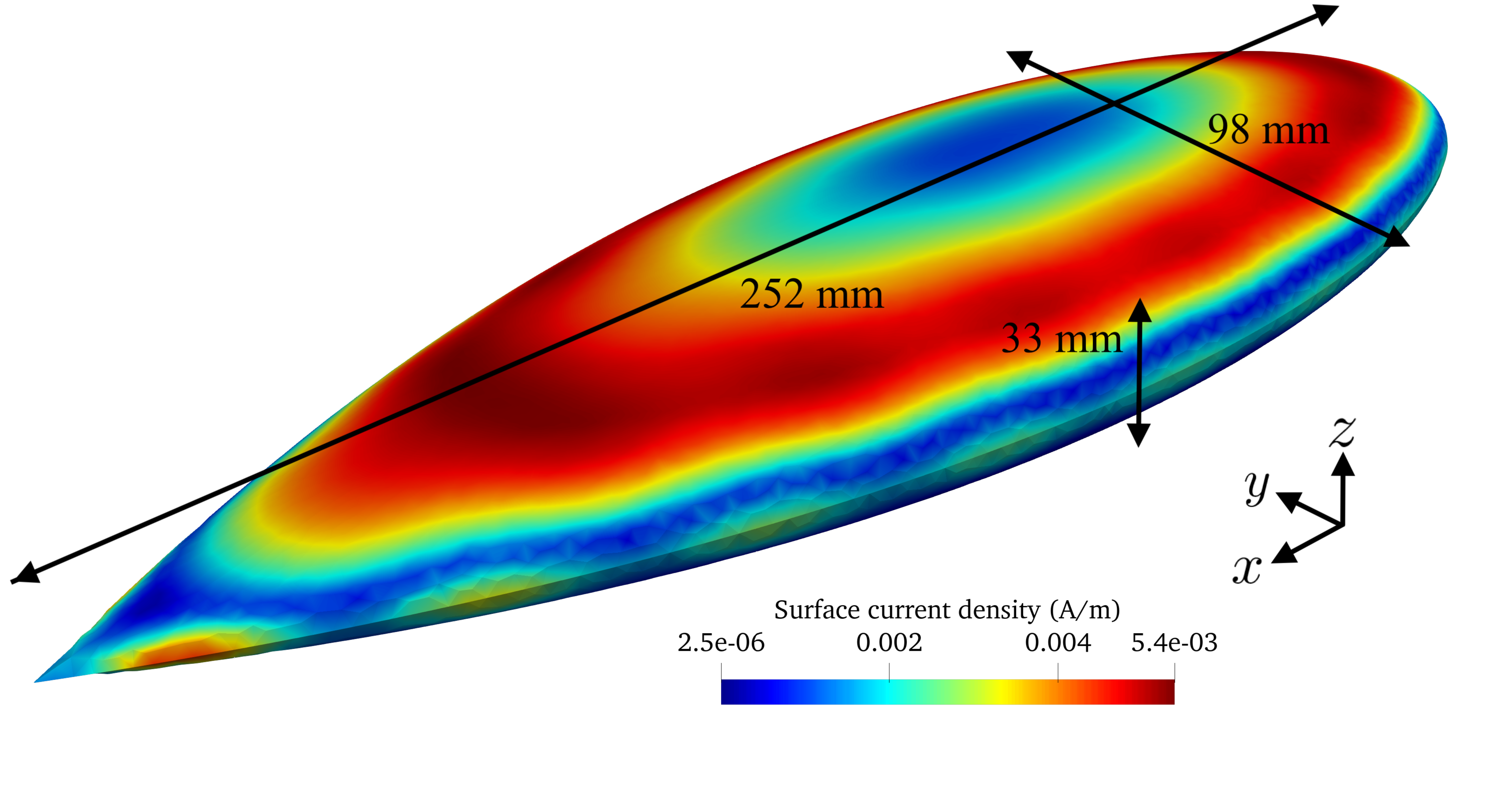}
  \caption{Geometry and electric surface current density at $10\,$GHz for the NASA almond in \secref{sec:results:alm}.}\label{fig:alm:J}
\end{figure}
\begin{figure}[t]
  \includegraphics[width=\linewidth]{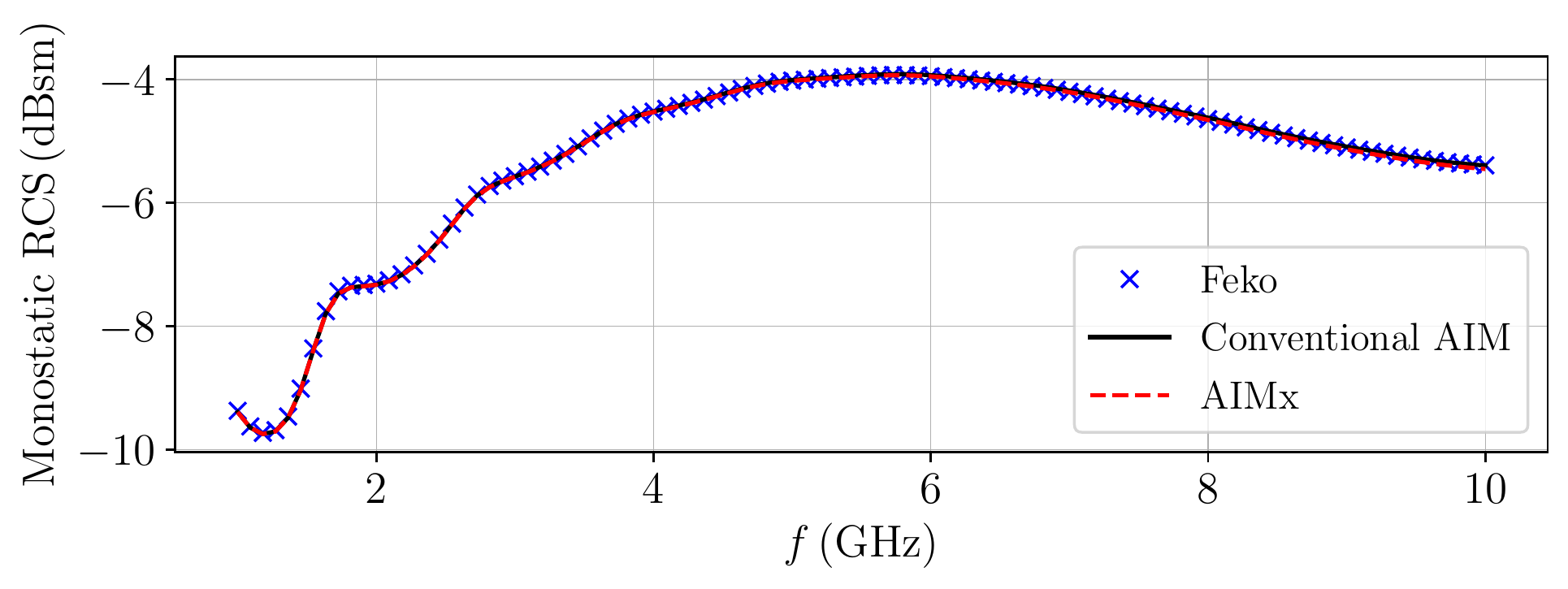}\label{fig:alm:mrcs}\\
  \includegraphics[width=\linewidth]{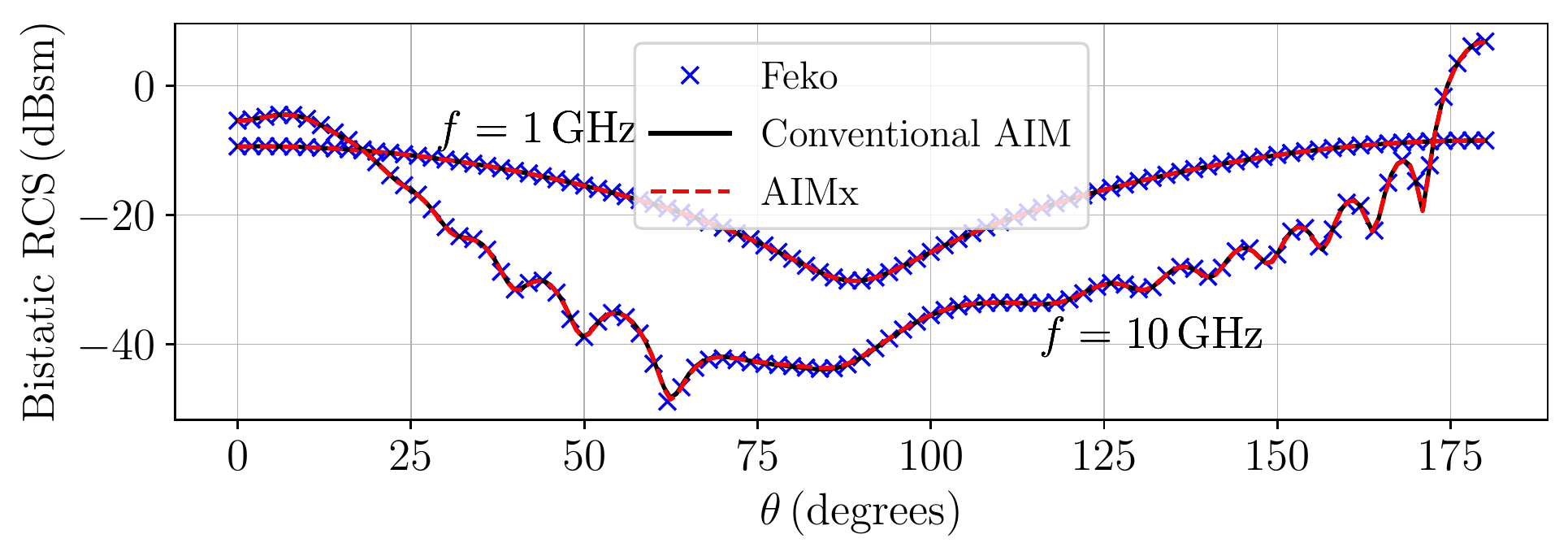}
  \caption{Accuracy validation for the NASA almond in \secref{sec:results:alm}. Top panel: monostatic RCS. Bottom panel: bistatic RCS at $1\,$GHz and $10\,$GHz.}\label{fig:alm:rcs}
\end{figure}
\begin{figure}[t]
  \includegraphics[width=\linewidth]{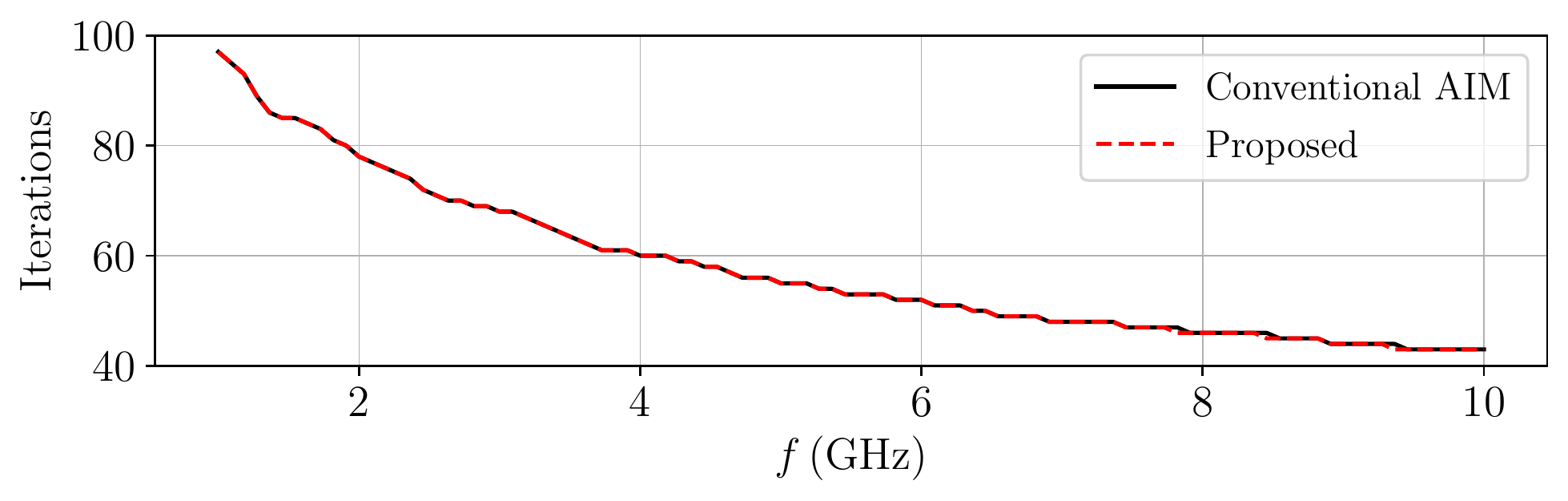}\\
  \includegraphics[width=\linewidth]{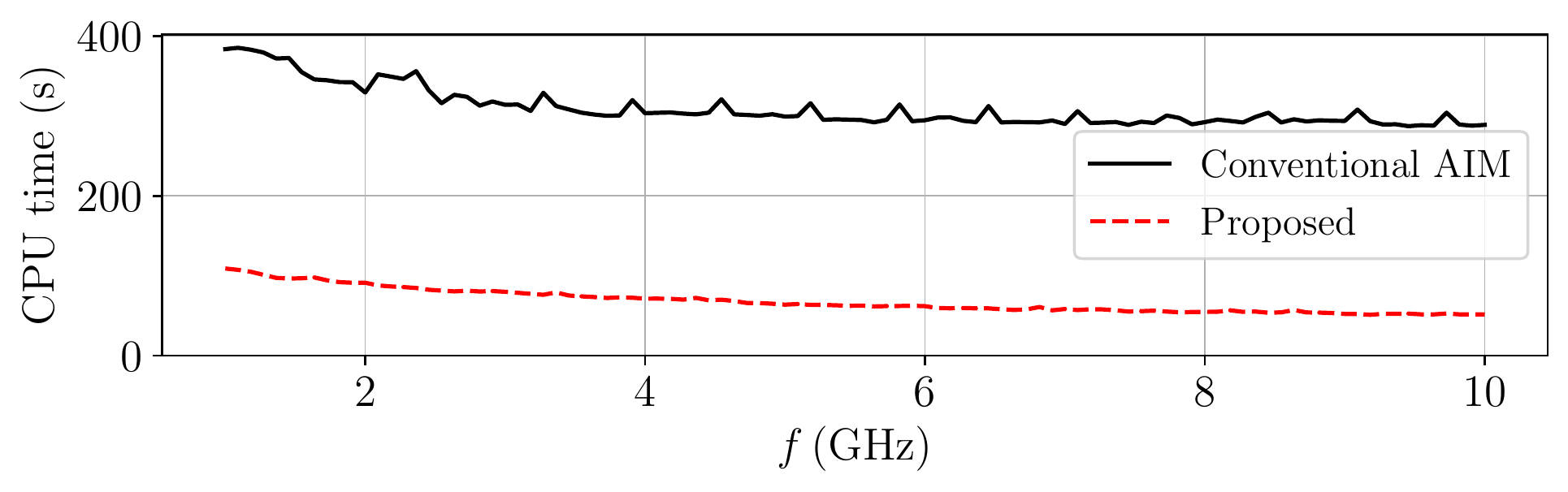}
  \caption{Performance comparison for the NASA almond in \secref{sec:results:alm}. Top panel: iterations required for convergence. Bottom panel: CPU time per frequency.}\label{fig:alm:prof}
\end{figure}

Next, we consider the NASA almond~\cite{nasaalmond} in free space, meshed with $26{,}796$ triangles and $40{,}194$ edges.
An AIM grid with ${126 \times 50 \times 17}$ points was used with ${n=2}$.
A plane wave with the electric field polarized along the $x$ axis is incident on the PEC structure, and the CFIE system~\eqref{eq:CFIEAIMx} was solved with a diagonal preconditioner.
The geometry and computed electric surface current density at $10\,$GHz are shown in \figref{fig:alm:J}.
The top pane of \figref{fig:alm:rcs} shows the monostatic RCS for $100$ frequency points spaced approximately $91\,$MHz apart, computed via AIMx, the conventional AIM, and the commercial method-of-moments solver Altair Feko.
Excellent agreement is observed across the entire frequency range.
The bottom panel of \figref{fig:alm:rcs} shows the bistatic RCS for the ${\phi = 0}$ cut, at $1\,$GHz and $10\,$GHz.
Again, the proposed method is in excellent agreement with the conventional AIM and with Feko.

The top panel of \figref{fig:alm:prof} shows that the proposed method requires the same number of iterations to converge as the conventional AIM across the entire frequency range, which reaffirms the effectiveness of the frequency-independent diagonal preconditioner compared to the full-wave version.
The bottom panel of \figref{fig:alm:prof} shows a significant reduction in the CPU time per frequency, compared to the conventional AIM.
The proposed method provided a total speed-up of $4.4\times$, and CPU times are reported in \tabref{tab:prof}.

\subsection{Planar Reflectarray}\label{sec:results:ref}

\begin{figure}[t]
  \centering
  \includegraphics[width=.83\linewidth]{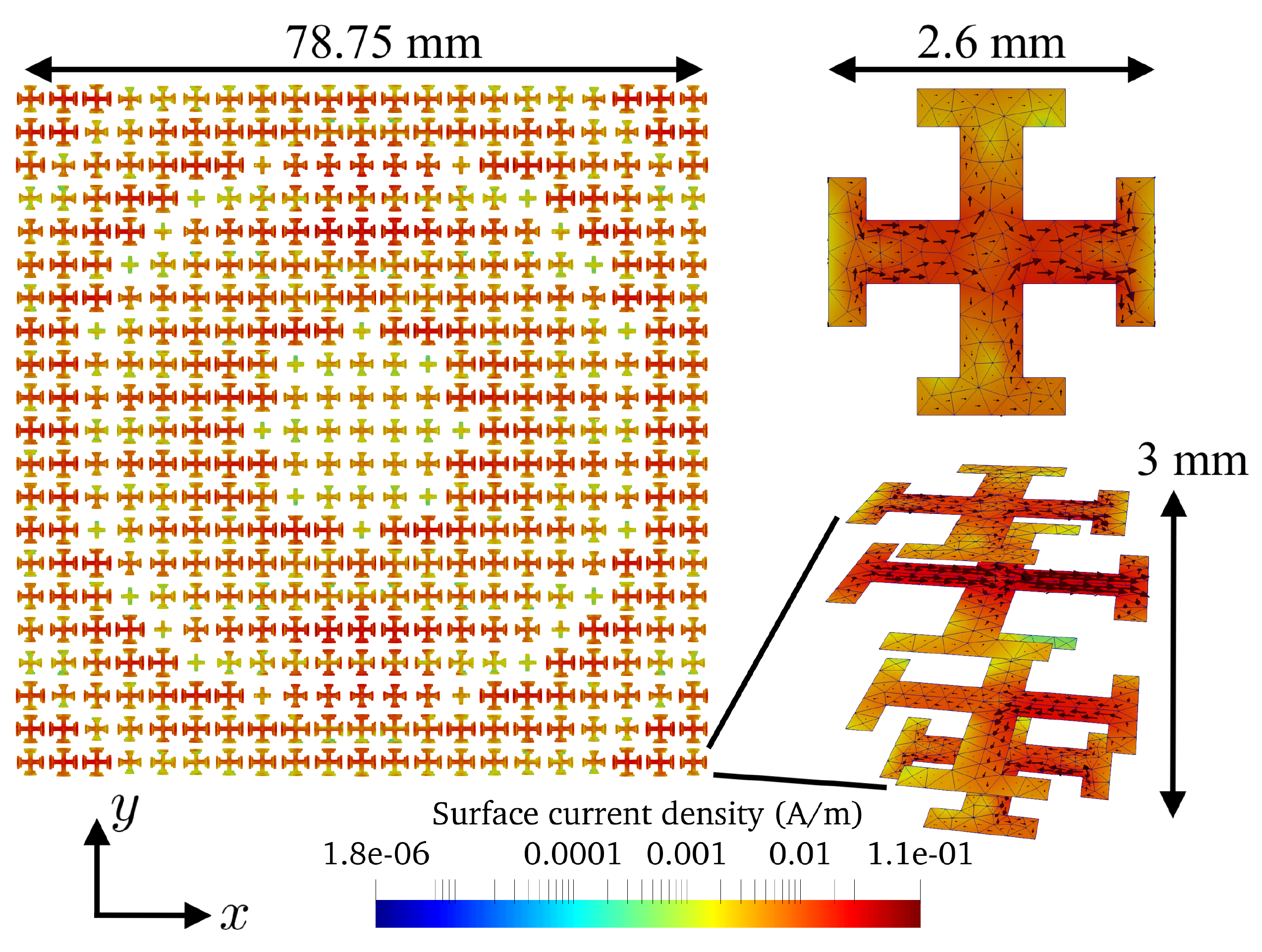}
  \caption{Geometry and electric surface current density at $25\,$GHz for the reflectarray in \secref{sec:results:ref}.}\label{fig:ref:J}
\end{figure}
\begin{figure}[t]
  \includegraphics[width=\linewidth]{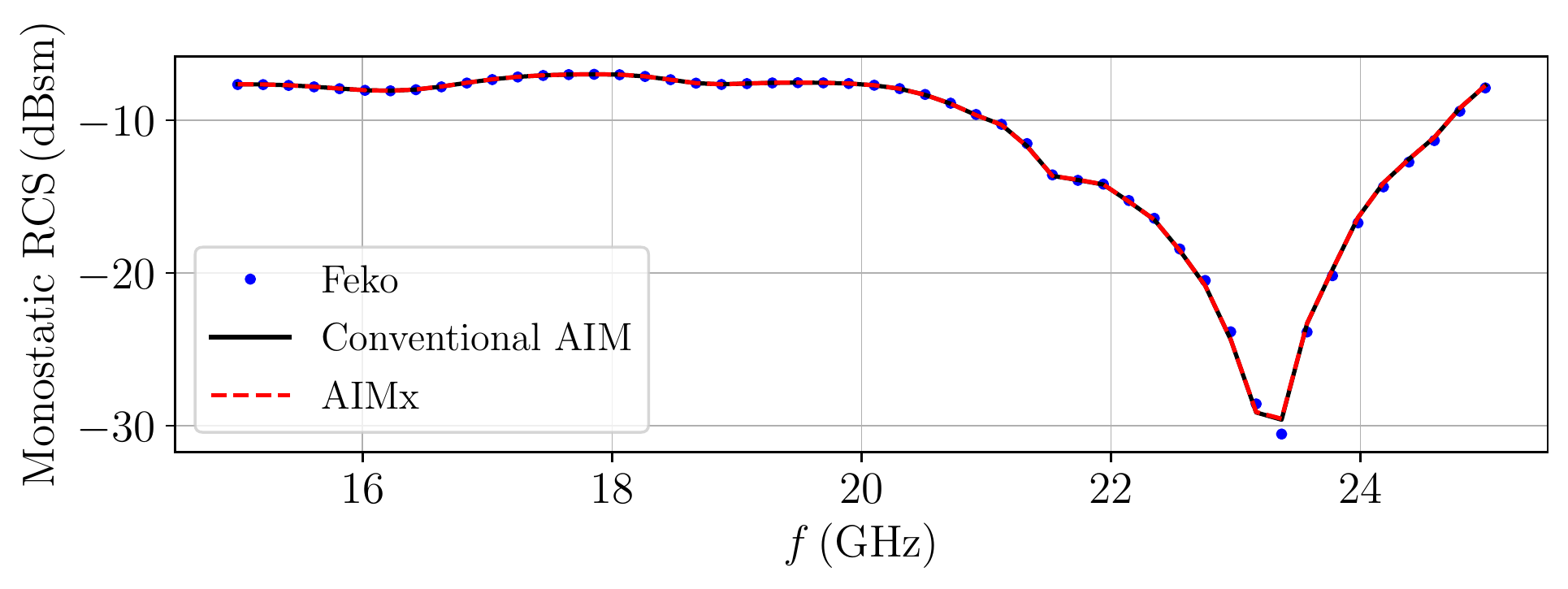}\\
  \includegraphics[width=\linewidth]{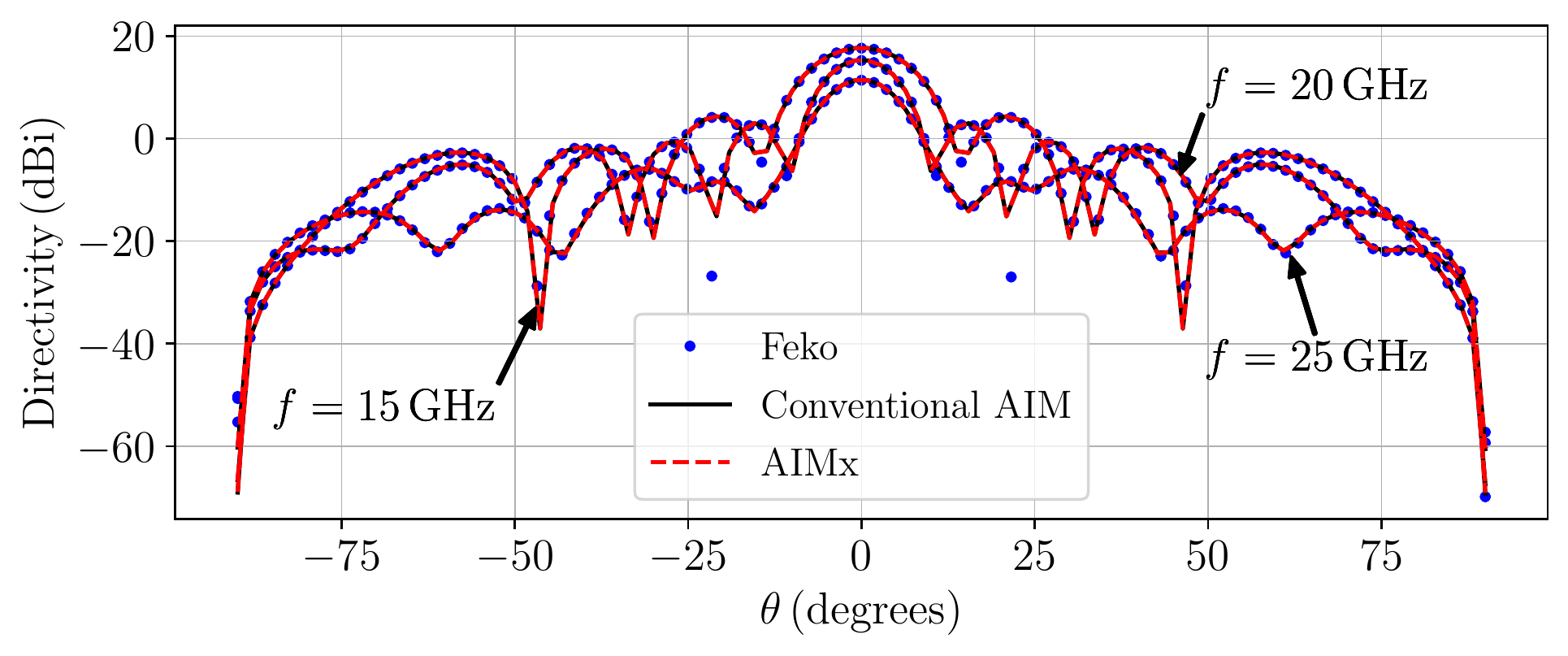}
  \caption{Accuracy validation for the reflectarray in \secref{sec:results:ref}. Top panel: monostatic RCS. Bottom panel: directivity at $15\,$GHz, $20\,$GHz and $25\,$GHz.}\label{fig:ref:rad}
\end{figure}
\begin{figure}[t]
  \includegraphics[width=\linewidth]{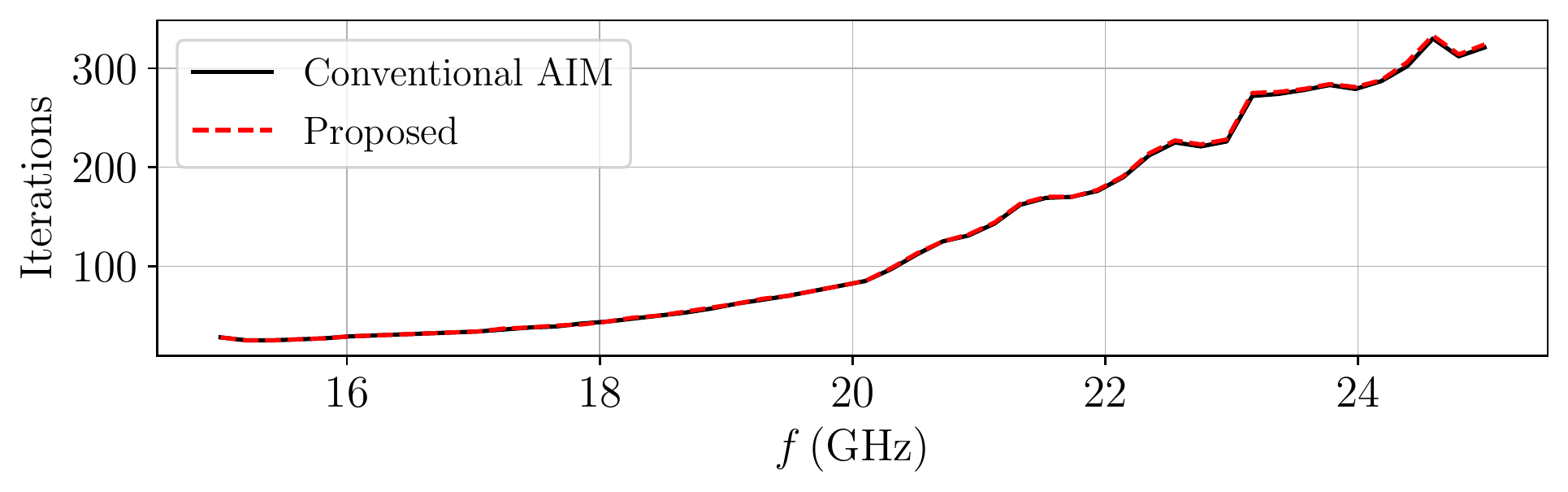}\\
  \includegraphics[width=\linewidth]{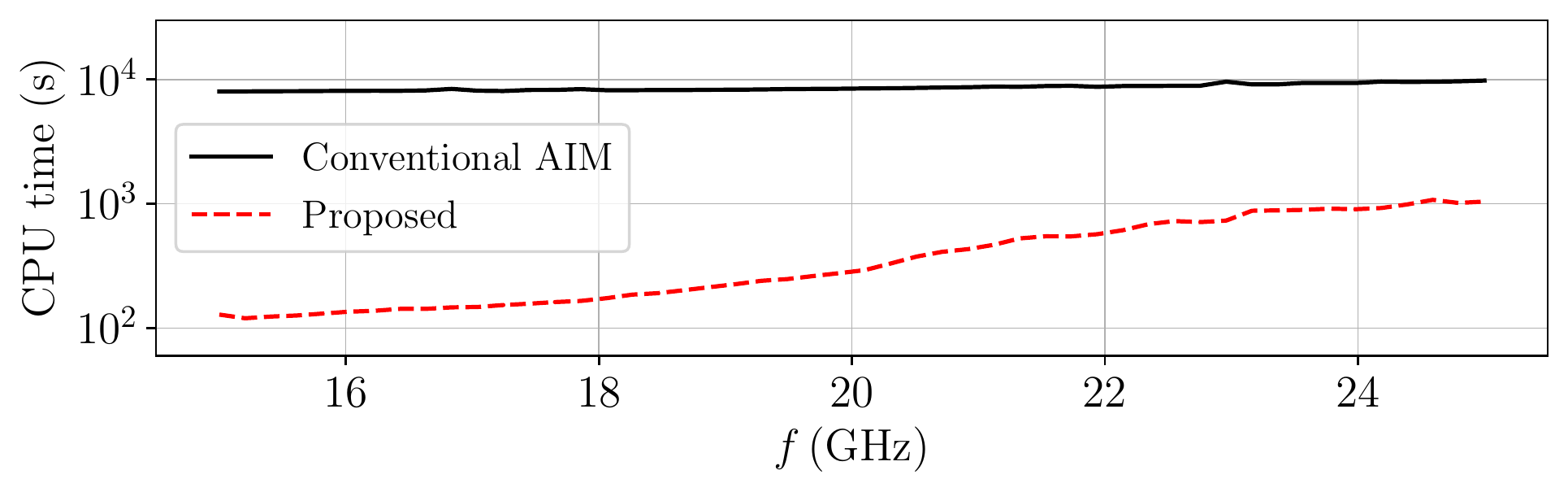}
  \caption{Performance comparison for the reflectarray in \secref{sec:results:ref}. Top panel: iterations required for convergence. Bottom panel: CPU time per frequency.}\label{fig:ref:prof}
\end{figure}

We next consider a more realistic test case involving a four-layer reflectarray in free space, with ${21 \times 21}$ planar PEC Jerusalem cross elements.
Details regarding the geometry of individual elements can be found in~\cite{utk_reflectarrayJC,hum_reflectarrayJC}.
The array is meshed with $254{,}736$ triangles and $324{,}420$ edges, and an AIM grid with ${158 \times 158 \times 7}$ points was used with ${n=3}$.
A plane wave with the electric field polarized along $x$ impinges on the array.
To robustly handle the intricate sub-wavelength features of the structure, the augmented EFIE (AEFIE)~\cite{aefie2} was used, which involves the same matrix operators as the EFIE~\eqref{eq:EFIEdis}, but takes the charge density as an additional unknown.
The constraint preconditioner~\cite{aefie2} involving self-terms of $\LmatA[\mathrm{s,NR}]$ and $\LmatPhi[\mathrm{s,NR}]$ was used.
The geometry and computed electric surface current density at $25\,$GHz are shown in \figref{fig:ref:J}.
The monostatic RCS for~$50$ evenly-spaced frequency points from~$15$ to~$25\,$GHz is shown in the top panel of \figref{fig:ref:rad}.
Over this range of frequencies, the electrical size of the structure ranges from $3.9$ to $6.6\,$wavelengths.
Additionally, the directivity for the ${\phi = 0}$ cut is shown in the bottom panel of \figref{fig:ref:rad} for $15$, $20$ and $25\,$GHz.
The RCS and directivity computed via the proposed method are both in excellent agreement with results from the conventional AIM, and with the commercial solver Feko.

The top panel of \figref{fig:ref:prof} shows that the preconditioner based on frequency-independent self-terms of $\LmatA[\mathrm{NR}]$ and $\LmatPhi[\mathrm{NR}]$ is as effective as the full-wave preconditioner proposed in~\cite{aefie2} across the entire frequency range considered.
Furthermore, the bottom panel of \figref{fig:ref:prof} shows the significantly improved efficieny of the proposed method compared to the conventional AIM.
At the lower end of the frequency range, fewer iterations are required in the solve step, and the CPU time in the conventional AIM is dominated by the near-region direct integration and precorrection steps.
At higher frequencies, the number of iterations increases, showing a commensurate increase in total CPU time in the proposed method, as shown in the bottom panel of \figref{fig:ref:prof}.
However, the overall cost of the conventional AIM is still dominated by the near-region computations.
The CPU times for both methods are reported in \tabref{tab:prof}.
It is important to note that besides the per-frequency time savings, the near-region direct integration step in the proposed method is nearly $3\times$ faster than in the conventional AIM.
This is because the static Green's function, $1/(4\pi r)$, is used as the kernel for near-region computations in the proposed method, which is faster to compute than the exponential function in the full-wave Green's function~\eqref{eq:hgf}, as required in the conventional AIM.
The proposed method reduces the total simulation time from over $5\,$ days to just $7.5\,$hours, which corresponds to an overall speed-up of $16.1\times$.

\subsection{Printed Dipole Array}\label{sec:results:dip}

\begin{figure}[t]
  \centering
  \includegraphics[width=.85\linewidth]{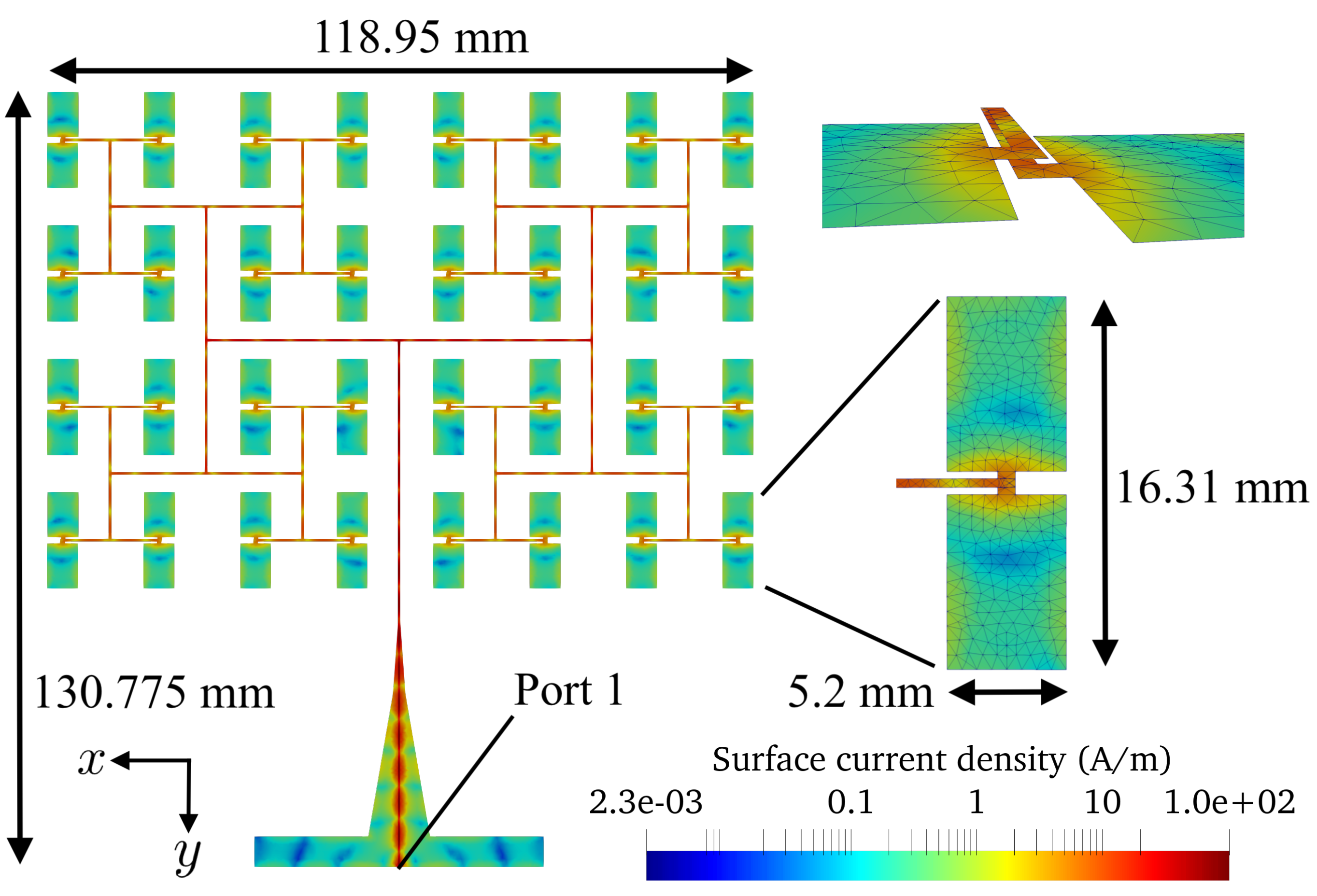}
  \caption{Geometry and electric surface current density at $20\,$GHz for the dipole array in \secref{sec:results:dip}.}\label{fig:dip:J}
\end{figure}
\begin{figure}[t]
  \includegraphics[width=\linewidth]{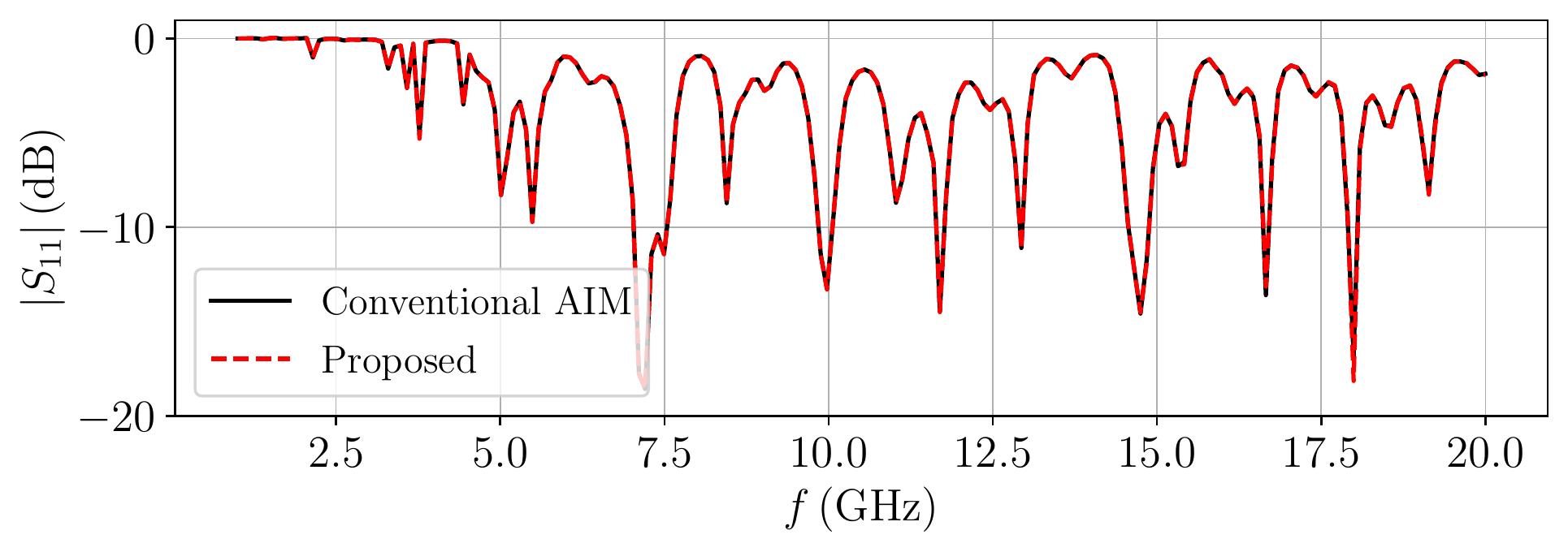}\\
  \includegraphics[width=\linewidth]{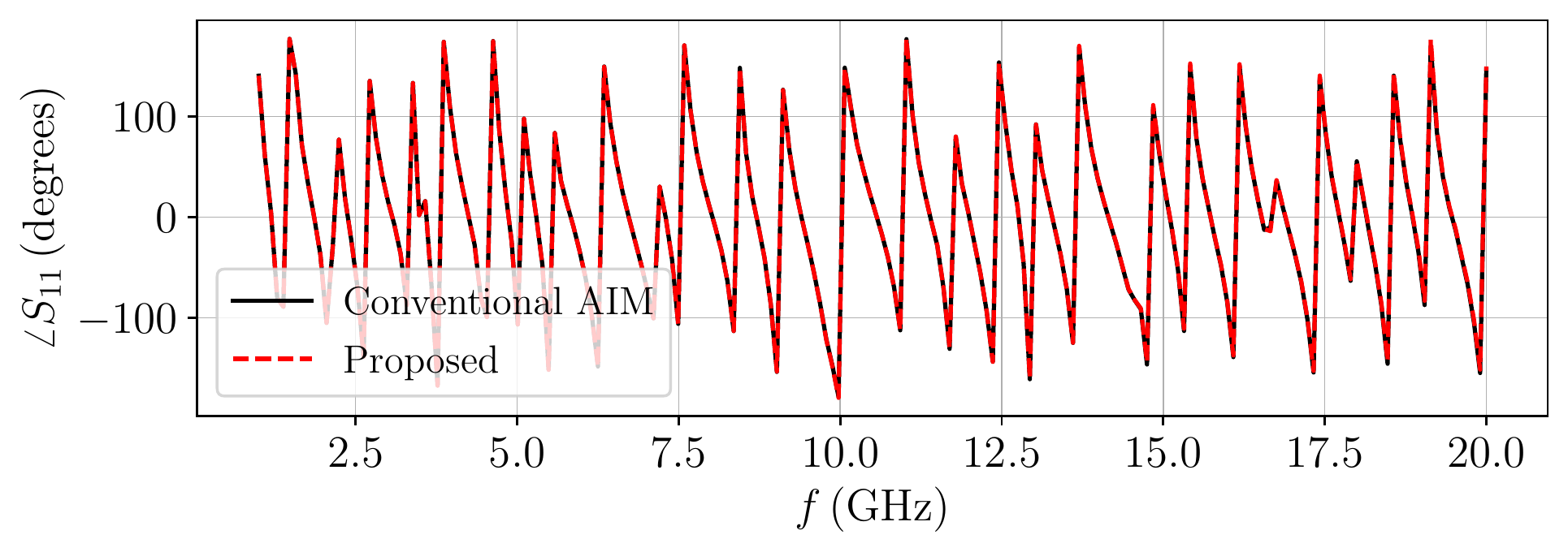}
  \caption{Scattering parameter validation for the dipole array in \secref{sec:results:dip}.
   }\label{fig:dip:S}
\end{figure}
\begin{figure}[t]
  \includegraphics[width=\linewidth]{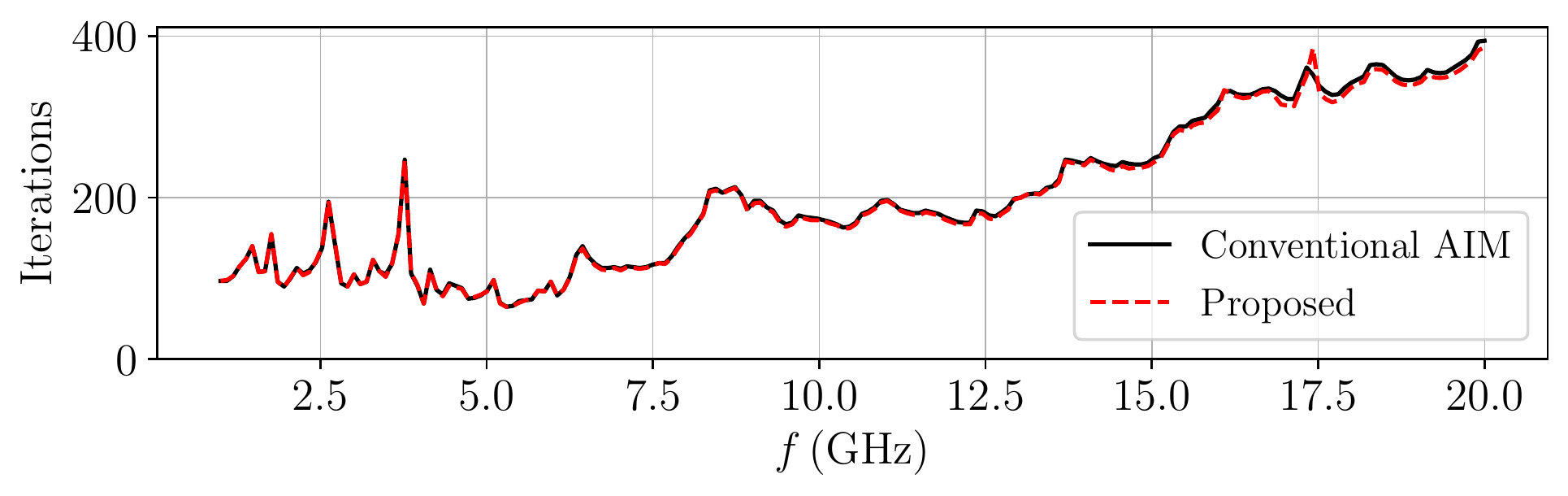}\\
  \includegraphics[width=\linewidth]{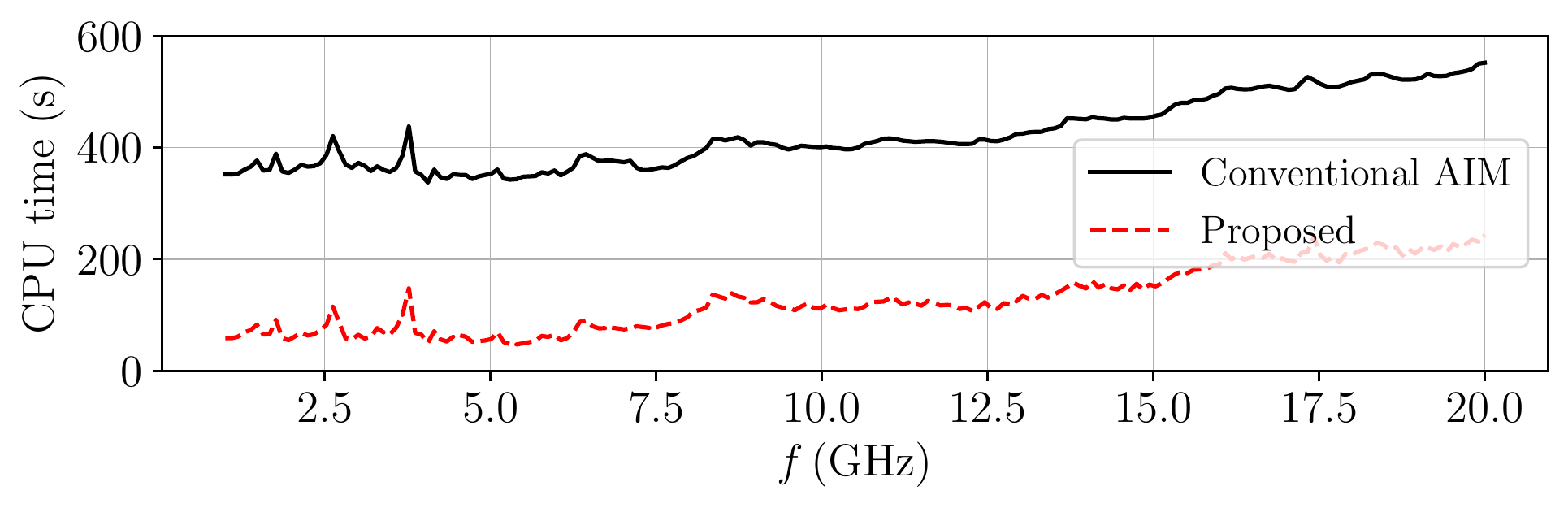}
  \caption{Performance comparison for the dipole array in \secref{sec:results:dip}. Top panel: iterations required for convergence. Bottom panel: CPU time per frequency.}\label{fig:dip:prof}
\end{figure}

The numerical examples considered so far involved structures in free space; the following examples concern structures embedded in layered media.
First, we consider a double-sided printed dipole array consisting of planar PEC elements fed by a balanced transmission line network.
The structure is shown in \figref{fig:dip:J}, and was inspired by~\cite{doubledipolearray}, which includes details of the geometry.
Each arm of the dipole elements is printed on opposide sides of a dielectric substrate with relative permittivity $2.17$, adjacent to a foam spacer backed by a ground plane.
The background medium configuration is shown in \tabref{tab:layers}.
The structure is meshed with $25{,}042$ triangles and $33{,}688$ edges, and an AIM grid with ${177 \times 177 \times 4}$ points was used with ${n=3}$.
The AEFIE formulation with the same constraint preconditioner as in \secref{sec:results:ref} was used, which now involves $\Grrpko[\phi]$ and $\Gdyadicrrpko[A]$ as kernels due to the layered background medium.
A delta-gap port~\cite{gibson} was employed to excite the structure, with the port location marked in \figref{fig:dip:J}, which also shows the computed electric surface current density at $20\,$GHz.
We consider~$200$ evenly-spaced frequency points from~$1$ to~$20\,$GHz, over which the structure spans electrical sizes between~$0.4$ and~$8$\,wavelengths.
The magnitude and phase of the scattering parameter $S_{11}$ are shown in the top and bottom panels of \figref{fig:dip:S}, respectively.
In both cases, the proposed method is in excellent agreement with the conventional AIM over the entire frequency range.
This demonstrates that the proposed method is also capable of accurately handling electrically large problems where the MGF is used to model layered background media.
The top panel of \figref{fig:dip:prof} again demonstrates the relative efficacy of the constraint preconditioner involving only frequency-independent self-term entries, while the bottom panel of \figref{fig:dip:prof} shows the per-frequency CPU time improvement of the proposed method over the conventional AIM.
\tabref{tab:prof} reports CPU times for the two methods, and shows that the near-region direct integration step in the proposed method is significantly faster than in the conventional AIM.
A total speed-up of $3.3\times$ was obtained with the proposed method.
In this case, the iterative solve time is a significant contributor to the total simulation time at each frequency, and is the limiting factor in the obtained speed-up.

\subsection{Stacked Inductor Array}\label{sec:results:ind}

\begin{figure}[t]
  \centering
  \includegraphics[width=.9\linewidth]{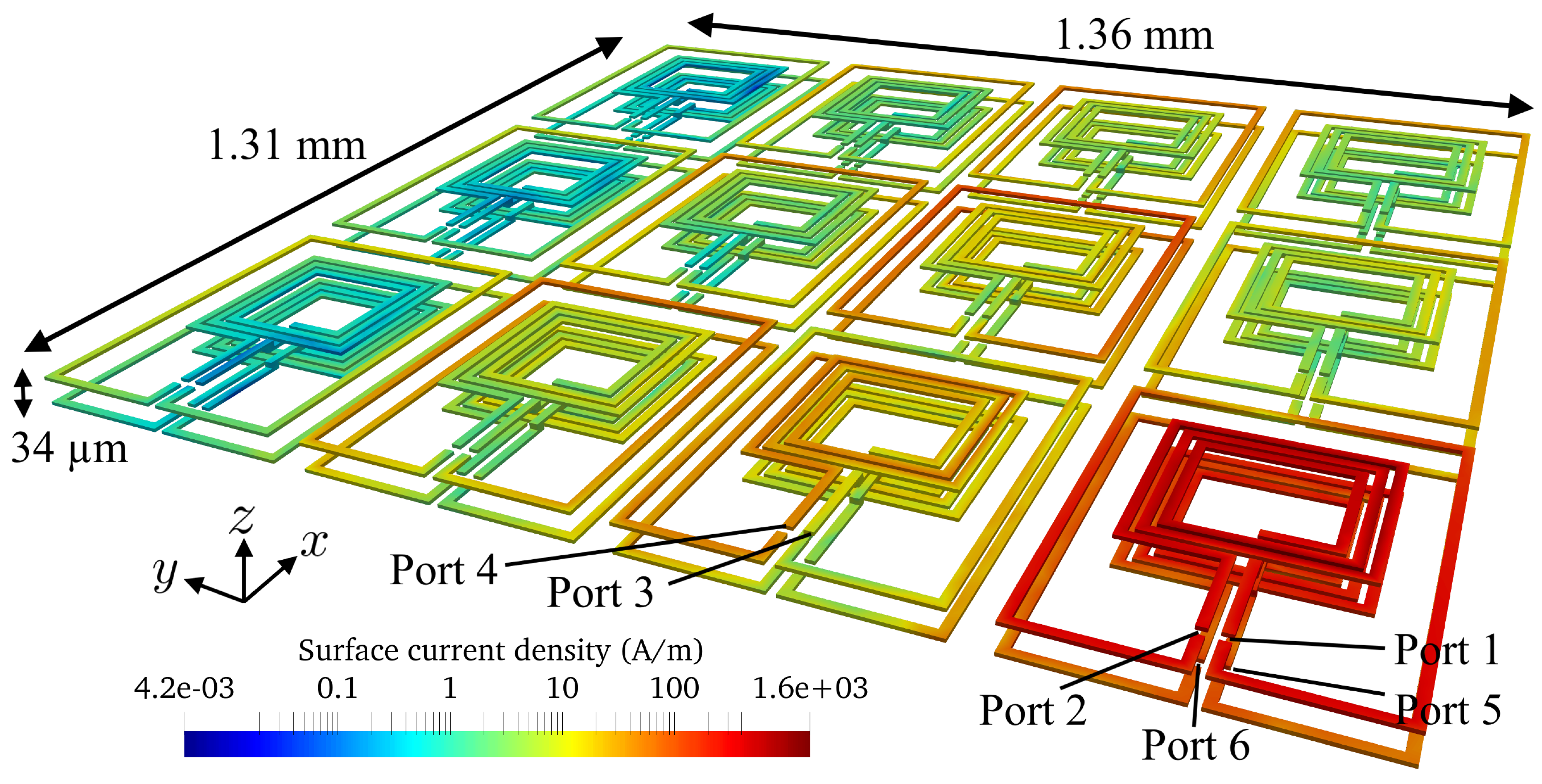}
  \caption{Geometry and electric surface current density at $60\,$GHz for the inductor array in \secref{sec:results:ind}.}\label{fig:ind:J}
\end{figure}
\begin{figure}[t]
  \includegraphics[width=\linewidth]{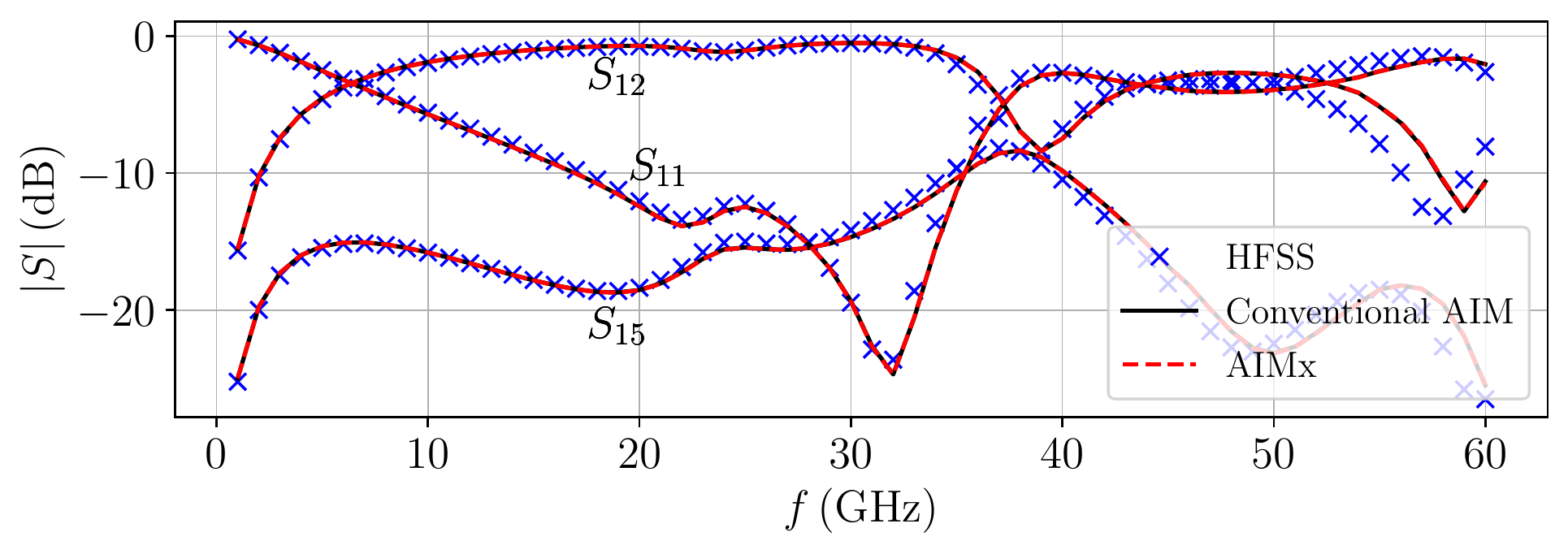}\\
  \includegraphics[width=\linewidth]{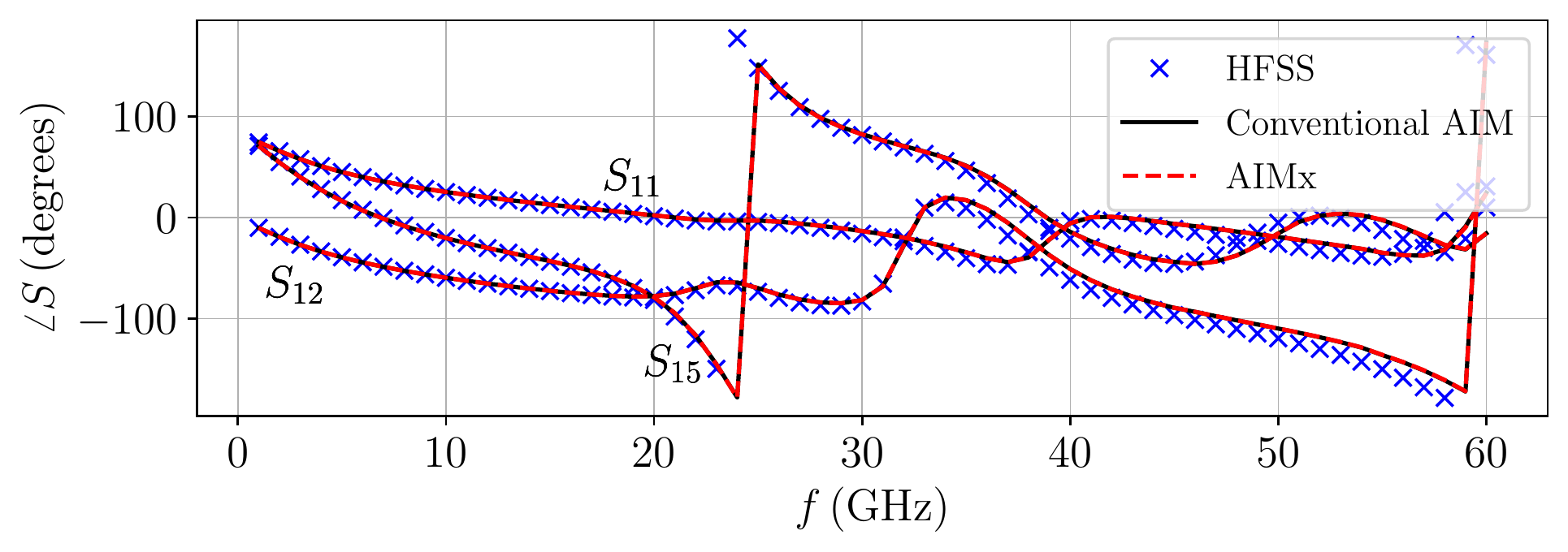}
  \caption{Scattering parameter validation for the inductor array in \secref{sec:results:ind}.
  }\label{fig:ind:S}
\end{figure}
\begin{figure}[t]
  \includegraphics[width=\linewidth]{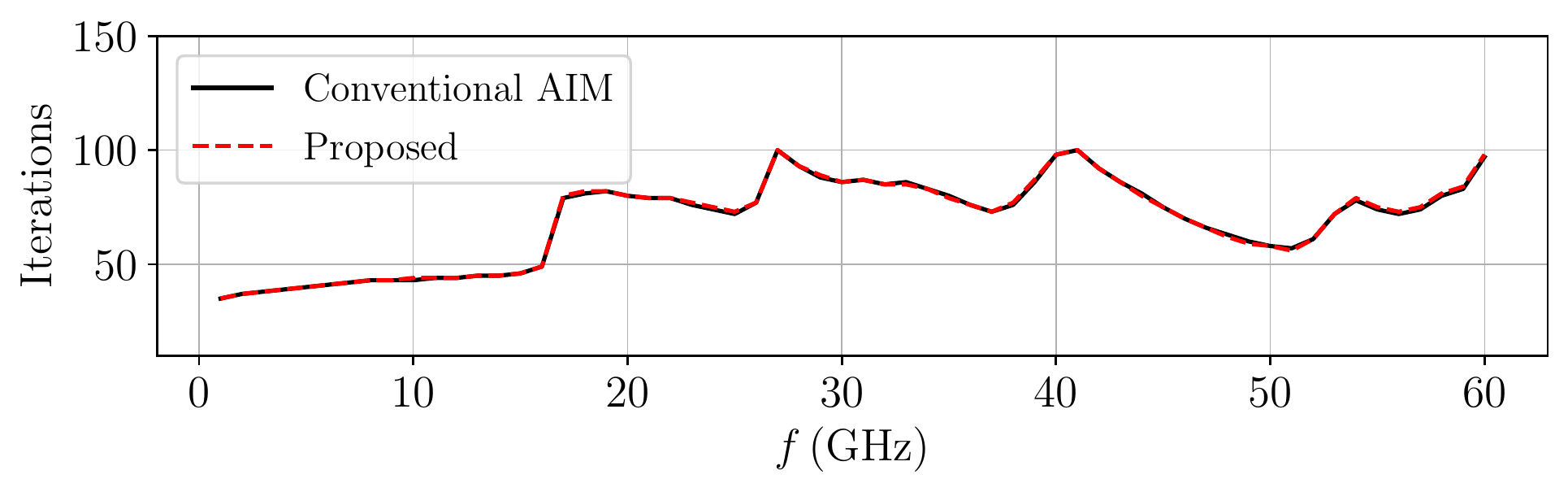}\\
  \includegraphics[width=\linewidth]{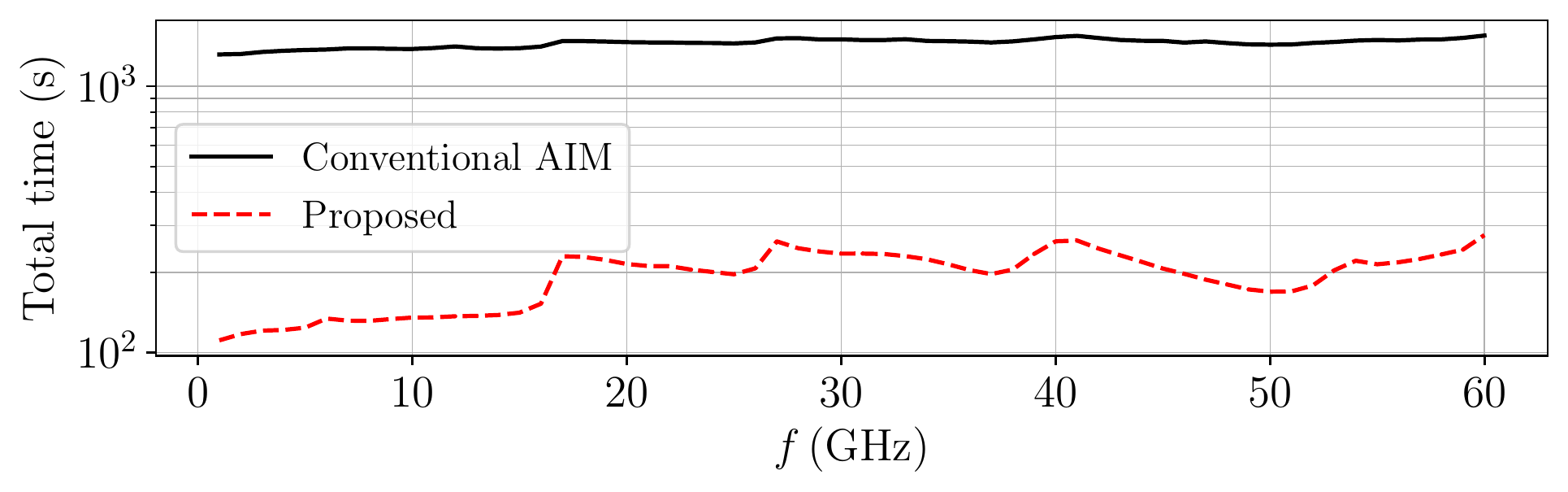}
  \caption{Performance comparison for the inductor array in \secref{sec:results:ind}. Top panel: iterations required for convergence. Bottom panel: CPU time per frequency.}\label{fig:ind:prof}
\end{figure}

As a final example, we consider an on-chip stacked inductor array~\cite{fastmaxwell} consisting of ${3 \times 4 \times 2}$ copper coils, shown in \figref{fig:ind:J}.
Each individual element is a $4\times$ scaled version of the geometry described in~\cite{EPEPS2017}.
The structure is embedded in a six-layer substrate with the configuration shown in \tabref{tab:layers}.
The first inductor layer starts at ${z=45}\,\mu$m, and the second at ${z=68}\,\mu$m, where ${z=0}$ is taken at the interface between the lowermost layer and the PEC backing.
A mesh with $42{,}396$ triangles and $63{,}594$ edges was generated, and an AIM grid with ${96 \times 100 \times 6}$ points was used with ${n=2}$.
The AEFIE formulation with the same constraint preconditioner as in the previous two examples was used, and the presence of skin effect was modeled with the surface impedance boundary condition (SIBC)~\cite{SIBC}.
The MGF involving $\Grrpko[\phi]$ and $\Gdyadicrrpko[A]$ was used to model the background medium.
A lumped port excitation model was used~\cite{gope,CPMT2019arxiv}, with port definitions shown in \figref{fig:ind:J}, along with the computed electric surface current density at $60\,$GHz.

The structure was simulated at~$60$ frequency points spaced~$1\,$GHz apart, between~$1\,$GHz and~$60\,$GHz.
The top and bottom panels of \figref{fig:ind:S} show, respectively, the magnitude and phase of~$S$ parameters corresponding to reflection, transmission, and coupling, for the proposed method, the conventional AIM, and the commercial finite element solver Ansys HFSS.
The proposed method is in excellent agreement in both magnitude and phase over the entire frequency range considered.
There are small deviations compared to Ansys HFSS at the highest frequencies, in both the proposed method and in the conventional AIM.
These deviations are most likely due to differences in how the lumped port excitation is defined.
Furthermore, the top panel of \figref{fig:ind:prof} confirms the effectiveness of the frequency-independent constraint preconditioner for all frequencies, while the bottom panel of \figref{fig:ind:prof} demonstrates the significant improvement in CPU time per frequency.
The near-region direct integration times are shown in \tabref{tab:prof}, and the proposed method leads to a speed-up of~$6.9\times$ in total, compared to the conventional AIM.

\section{Discussion}\label{sec:discussion}

Through various numerical examples, we demonstrated that the proposed method greatly speeds up frequency sweeps in SIE-based electromagnetic solvers, while maintaining good accuracy.
Specifically, we emphasize the following advantages arising from the proposed technique:
\begin{itemize}
  \item The cost-per-frequency of the matrix-fill and precorrection steps is eliminated.
  Instead, these steps are performed only once at the beginning of the frequency sweep.
  For $N$ frequency points, the CPU time in the proposed method is approximately ${T_\mathrm{AIMx} = T_\mathrm{s,AIMx} + N\,T_\mathrm{d, AIMx}}$, where $T_\mathrm{s,AIMx}$ is the one-time cost of direct integration and precorrections, and $T_\mathrm{d,AIMx}$ is the average per-frequency cost, which mostly comprises the iterative solve time.
  In the conventional AIM, the total simulation time may be written as ${T_\mathrm{AIM} = N\,T_\mathrm{d, AIM}}$, where $T_\mathrm{d,AIMx}$ is the average per-frequency cost.
  For sufficiently large $N$, ${T_\mathrm{s,AIMx} \ll N\,T_\mathrm{d, AIMx}}$.
  Past this point, the proposed method yields a maximum relative speed-up of ${T_\mathrm{d, AIMx}}/{T_\mathrm{d, AIM}}$ per frequency point.

  \item Direct integration in the near-region is greatly simplified, since only frequency-independent kernels are involved.
  Therefore, for layered media, expensive Sommerfeld integrals or approximation techniques such as the DCIM are avoided in the direct integration phase.
  \item For complex geometries, there is usually a tradeoff between mesh quality and CPU time, since a high quality mesh may require more elements.
  The cost of using high-quality meshes is significantly alleviated with the proposed method, since direct integration and precorrections are performed only once per frequency sweep.
  \item One may precompute and store to disk the frequency-independent near-region matrix for a given structure.
  The matrix could be loaded and reused whenever needed, for example when further refinement is needed near a resonance, or when the frequency range must be extended.
\end{itemize}

% ===================================================================

\section{Conclusion}\label{sec:conclusion}
A novel technique is proposed for the accelerated electromagnetic modeling of large problems in homogeneous and layered media, based on an extended adaptive integral method (AIMx).
The kernels of the SIE operators are decomposed into frequency-independent and frequency-dependent parts, such that the singularity of the kernel is contained entirely in the frequency-independent part.
The frequency-dependent part can then be captured with good accuracy via FFTs on an auxiliary grid, even for nearby source and observation points, as shown with a detailed error analysis.
% Therefore, all matrix-vector products involving the frequency-dependent terms are accelerated with FFTs in both the near- and far-region.
Therefore, near-region direct integration is required only for the frequency-independent term, which can be performed once in advance of a frequency sweep.
This leads to significantly faster sweeps than possible with conventional acceleration methods.
The efficiency of the proposed AIMx is demonstrated through diverse numerical examples, leading to total speed-ups ranging from $3\times$ to $16\times$.

\section*{Acknowledgment}

The authors would like to thank the anonymous reviewers for their thoughtful and constructive feedback.

\ifCLASSOPTIONcaptionsoff
  \newpage
\fi


%\clearpage

\bibliographystyle{IEEEtran}
%\bibliographystyle{ieeetr}
\bibliography{IEEEabrv,./bibliography}

\begin{IEEEbiography}[{\includegraphics[width=1in,height=1.25in,clip,keepaspectratio]{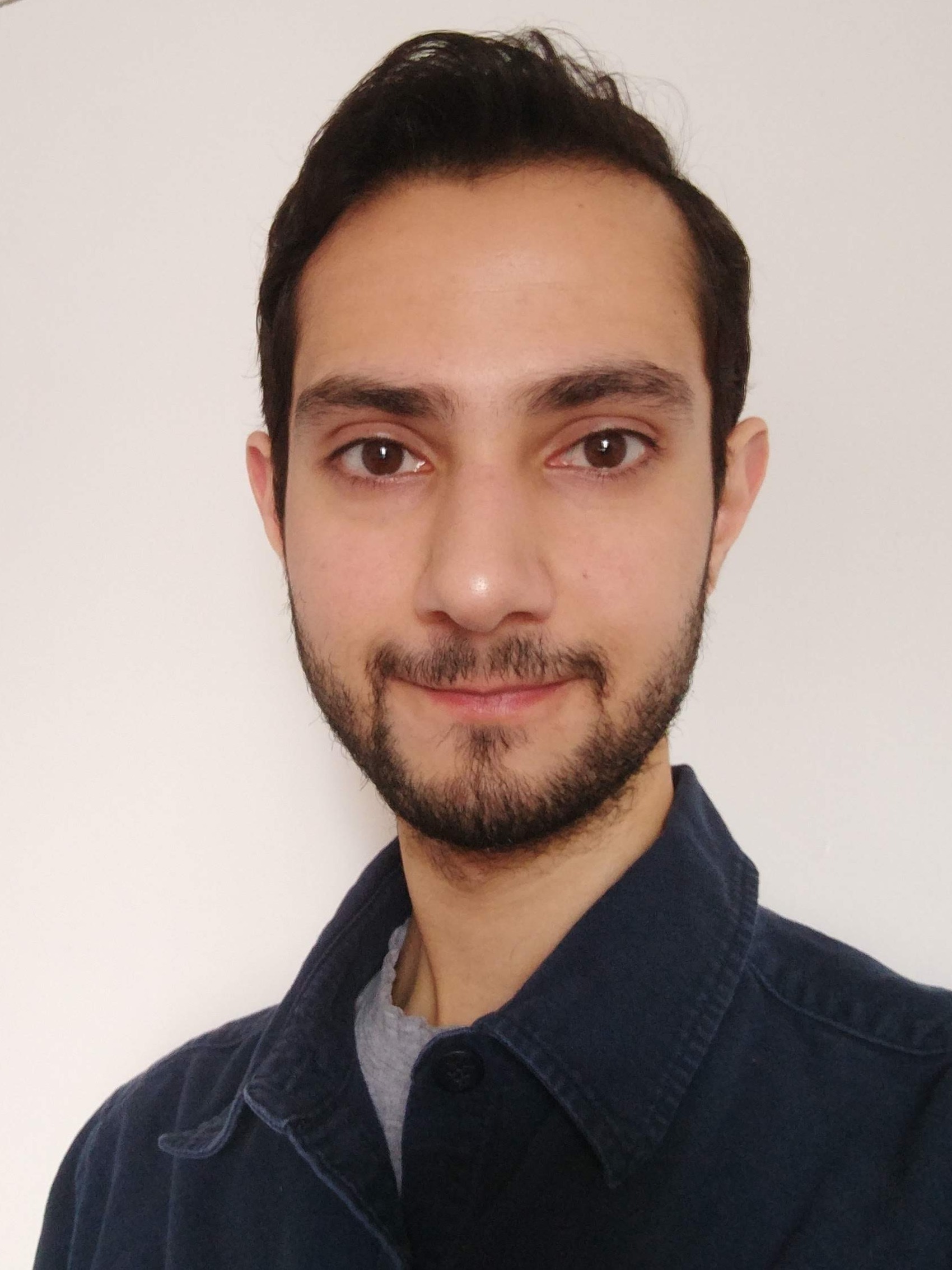}}]{Shashwat Sharma} (S'18)
received the B.A.Sc. degree in engineering physics and the M.A.Sc. degree in electrical engineering from the University of Toronto (U of T), Canada, in 2014 and 2016, respectively. From September 2016 to March 2017 he worked as a computational science research intern at Autodesk, Toronto. Since 2017 he has been working towards the Ph.D. degreee in electrical engineering at U of T. His research focuses on computational electromagnetics, with an emphasis on fast and robust integral equation methods for multiscale electromagnetic modeling. His interests include all aspects of mathematical modeling and scientific computing applied to electromagnetics.

Mr.~Sharma placed second at the CNC/USNC-URSI Student Paper Competition of the 2020 IEEE International Symposium on Antennas and Propagation and North American Radio Science Meeting (AP-S/URSI), and was a finalist for the Best Student Paper Award at the IEEE Conference on Electrical
Performance of Electronic Packaging and Systems (2018). He also received an honorable mention for his contributions to the AP-S/URSI symposia in 2019 and 2020, and a TICRA-EurAAP Grant at the European Conference on Antennas and Propagation (2021). He is currently serving as the vice chair for the U of T student chapter of the IEEE Antennas and Propagation Society.
\end{IEEEbiography}

\begin{IEEEbiography}[{\includegraphics[width=1in,height=1.25in,clip,keepaspectratio]{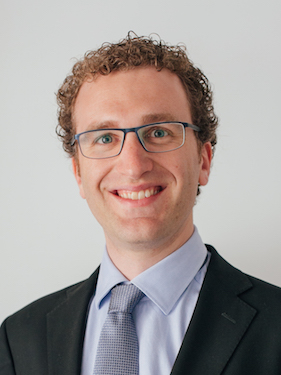}}]{Piero
Triverio} (S'06 -- M'09 -- SM'16)
received the Ph.D. degree in Electronic Engineering from Politecnico di
Torino, Italy, in 2009. He is an Associate Professor in The Edward S.
Rogers Sr. Department of Electrical \& Computer Engineering (ECE) at the
University of Toronto, and in the Institute of Biomaterials and
Biomedical Engineering (IBBME). He holds the Canada Research Chair in
Computational Electromagnetics. His research interests include signal
integrity, computational electromagnetism, model order reduction, and
computational fluid dynamics applied to cardiovascular diseases.

Prof.~Triverio received the Best Paper Award of the IEEE Transactions on
Advanced Packaging (2007), the EuMIC Young Engineer Prize (2010),  the
Connaught New Researcher Award (2013), and the Ontario Early Researcher
Award (2016). From 2013 to 2018, Triverio held the Canada Research Chair
in Modeling of Electrical Interconnects. Triverio and his students were
awarded the Best Paper Award of the IEEE Conference on Electrical
Performance of Electronic Packaging and Systems (2008, 2017), and
several Best Student Paper Awards at international symposia. He serves
as an Associate Editor for the IEEE TRANSACTIONS ON COMPONENTS,
PACKAGING AND MANUFACTURING TECHNOLOGY. He is a member of the Technical
Program Committee of the IEEE Workshop on Signal and Power Integrity,
and of the IEEE Conference on Electrical Performance of Electronic
Packaging and Systems.
\end{IEEEbiography}

\end{document}